\documentclass[english,aps,preprintnumbers,amsmath,amssymb,nofootinbib,twocolumn,superscriptaddress]{revtex4}

\usepackage[latin1]{inputenc}
\usepackage{graphicx}

\usepackage{color}
\definecolor{darkgreen}{rgb}{0.00, 0.60, 0.00}

\definecolor{turquois}{rgb}{0.00, 0.70, 0.70}

\usepackage{ulem}

\usepackage{bm}
\usepackage{amsmath}

\usepackage{dsfont}

\newcommand{\Tr}{\mathrm{Tr}}
\newcommand{\mf}{m_{\rm f}}
\newcommand{\mfb}{\bar{m}_{\rm f}}
\newcommand{\mfbT}{\bar{m}_{{\rm f},T}}

\newcommand{\be}{\begin{eqnarray}}
\newcommand{\ee}{\end{eqnarray}}

\newcommand{\nn}{\nonumber }

\newcommand{\Tc}{T_{\text{cr}}}

\newcommand{\xsb}{$\chi$SB}
\newcommand{\Nf}{N_{\text{f}}}

\newcommand{\pat}{\partial_t}
\newcommand{\Eqref}[1]{Eq.~\eqref{#1}}

\newcommand{\Polylog}[1]{\mathrm{Li}_{#1}}

\newcommand{\fslash}{\hspace{-0.1ex} \slash }
\def\slash#1{\setbox0=\hbox{$#1$}               % set a box for #1
   \dimen0=\wd0                                 % and get its size
   \setbox1=\hbox{/} \dimen1=\wd1               % get size of /
   \ifdim\dimen0>\dimen1                        % #1 is bigger
      \rlap{\hbox to \dimen0{\hfil/\hfil}}      % so center / in box
      #1                                        % and print #1
   \else                            
      \rlap{\hbox to \dimen1{\hfil$#1$\hfil}}   % so center #1
      /                                         % and print /
   \fi}                                         %

\newcommand{\psibar}{\bar{\psi}}
\newcommand{\spinor}{\psi_{j}}
\newcommand{\barspinor}{\bar{\psi}_{j}}
\newcommand{\altbarspinor}{\bar{\psi}_{i}}
\newcommand{\altspinor}{\psi_{i}}

\newcommand{\wavpsi}{Z_{\psi}}

\usepackage{babel}
\makeatother

\begin{document}

\title{Many-flavor Phase Diagram of the $(2+1)d$ Gross-Neveu Model at Finite Temperature}
%\date{\today}                                           

\author{Daniel D.~Scherer}
\affiliation{Institut f\"ur Theoretische Physik, Universit\"at Leipzig, D-04103 Leipzig, Germany}
\affiliation{Theoretisch-Physikalisches Institut, Friedrich-Schiller-Universit\"at Jena, Max-Wien-Platz 1, D-07743 Jena, Germany}

\author{Jens Braun} 
\affiliation{Theoretisch-Physikalisches Institut, Friedrich-Schiller-Universit\"at Jena, Max-Wien-Platz 1, D-07743 Jena, Germany} 
\affiliation{Institut f\"ur Kernphysik
  (Theoriezentrum), Technische Universit\"at Darmstadt, D-64289 Darmstadt, Germany} 

\author{Holger Gies}
\affiliation{Theoretisch-Physikalisches Institut, Friedrich-Schiller-Universit\"at Jena, Max-Wien-Platz 1, D-07743 Jena, Germany}

\begin{abstract}
  We study the phase diagram of the Gross-Neveu model in $d=2+1$
  space-time dimensions in the plane spanned by temperature and the
  number of massless fermion flavors. We use a functional
  renormalization group approach based on a nonperturbative derivative
  expansion that accounts for fermionic as well as composite bosonic
  fluctuations. We map out the phase boundary separating the ordered
  massive low-temperature phase from the disordered high-temperature
  phase. The phases are separated by a second-order phase transition
  in the $2d$ Ising universality class. We determine the size of the
  Ginzburg region and show that it scales to zero for large $\Nf$
  following a powerlaw, in agreement with large-$\Nf$ and
    lattice studies. We also study the regimes of local order above
  as well as the classical regime below the critical temperature.
\end{abstract}

\maketitle

%----------------------------------------------------------------
\section{Introduction}
%----------------------------------------------------------------

The Gross-Neveu model \cite{Gross:1974jv} in
$(2+1)$-dimensions exemplifies many of the intriguing properties of
interacting relativistic fermion systems in a planar spacetime. From
a theory perspective, it is an ideal laboratory to study the interplay
of initially massless fermions with dynamically generated collective
excitations. From the viewpoint of applications, the study of effective
quantum field theories of low-dimensional Dirac fermions has become a rapidly evolving
field in the context of condensed-matter systems like graphene \cite{Novoselov:2005kj}, topological
band \cite{Hasan:2007,Hasan:2010xy,Qi:2011} and Mott insulators
\cite{Raghu}, as well as the normal-state behavior of high-temperature
superconductors \cite{SC1,SC2}.

The Gross-Neveu model in $(2+1)$-dimensions exhibits a quantum phase
transition to a phase of broken (discrete) chiral symmetry as a
function of the fermionic interaction strength. Despite being
perturbatively nonrenormalizable, it provides a paradigmatic example
for an asymptotically safe theory \cite{Weinberg:1976xy} that is
ultraviolet (UV) complete due to the existence of an interacting non-Gau\ss ian
renormalization group (RG) fixed point
\cite{Weinberg:1996kw}.
This fact has been proven to all orders in a $1/\Nf$ expansion
\cite{Gawedzki:1985ed,Rosenstein:1988pt,deCalan:1991km} and is
supported by lattice simulations \cite{Hands:1992ck,Hands:1992be,Karkkainen:1993ef}
as well as functional RG studies \cite{Braun:2010tt}. The quantum
phase transition associated with the RG fixed point has been studied
with a variety of techniques
\cite{Gat:1991bf,Gracey:1993kc,Vasiliev:1992wr,Karkkainen:1993ef,Braun:2010tt,Christofi:2006zt,Focht:1995ie,Reisz:1997ic,Babaev:1999in,Rosa:2000ju,Hofling:2002hj,Sonoda:2011qd}. 

Asymptotic safety makes the Gross-Neveu model in $(2+1)$-dimensions a
complete theory which is valid on all scales. It also ensures a
``perfect'' predictive power: once, a single (scale) parameter is
fixed, each and every physical observable can be predicted
absolutely. This statement extends to all dimensions $2\leq d < 4$,
with the non-Gau\ss ian fixed point merging with the Gau\ss ian fixed
point for $d\to 2$ \cite{Braun:2010tt}. 

From the viewpoint of critical phenomena, the properties of the
UV fixed-point control the universal features of the quantum phase
transition. Quantitatively, this statement extends to the long-range
scaling behavior of the chiral order parameter and/or the correlation
length near the phase transition which are both governed by the critical
exponents of the RG fixed point. Also for these critical exponents,
the quantitative agreement among the various methods is very
satisfactory, see
e.g. \cite{Karkkainen:1993ef,Christofi:2006zt,Gracey:1993kc,Braun:2010tt}. 

Whereas the quantum phase transition in the Gross-Neveu model in
$2<d<4$ dimensions is governed by the coupling strength as a control
parameter, many low-dimensional fermion systems can also develop
interesting features or even phase transitions as a function of the
flavor number $\Nf$, i.e., the number of interacting fermion
species. {For instance, convincing evidence has been
  collected for the fact that the $(2+1)$-dimensional Thirring model
  features chiral symmetry breaking (\xsb) only for flavor numbers
  below a critical value $\Nf<N_{\text{f,cr}}$
  \cite{Gomes:1991,Kondo:1995np,Christofi:2007ye,Gies:2010st}. A
  similar behavior is found for fermion systems with gauge
  interactions such as QED$_{(2+1)}$ \cite{Pisarski:1984dj}
 or even QCD$_{(3+1)}$ \cite{Caswell:1974gg}.  
 The latter class of many-flavor theories is
currently actively discussed in the context of New Physics scenarios
beyond the standard-model of particle physics \cite{Weinberg:1979bn}.} 

In all these cases, non-analytic features as a function of the flavor
number $\Nf$ occur, limiting the validity region of the large-$\Nf$
expansion. {The occurrence of a critical flavor number in these models
  arises from a competition between symmetry stabilizing and
  destabilizing fluctuations.} In this respect, the
$(2+1)$-dimensional Gross-Neveu model behaves less dramatically as a
function of $\Nf$.  Qualitatively, this can be understood from the
{fact that only one relevant bosonic degree of freedom corresponding
  to a ``radial'' $\mathds{Z}_2$ mode cannot counterbalance the
  destabilizing fermion fluctuations at strong coupling.}  Still, the
Gross-Neveu model is not completely structureless as a function of
flavor number, as is indicated by several observations: first, in the
limit $\Nf\to 0$, fermion interactions are switched off leaving us
with a ${\mathbb Z}_2$-symmetric universality class of (collective)
scalar degrees of freedom and a corresponding Ising-type phase
transition. This implies that the Gross-Neveu universality class has
to merge with the $3d$ Ising universality class in the limit of small
flavor number, approaching a Wilson-Fisher fixed point. From the
large-$\Nf$ expansion, it is already visible that this merger requires
a non-monotonic behavior in the universal quantities. Second, the
large-$\Nf$ limit (by definition) merges exactly with mean-field
theory which therefore must also hold for the scaling of observables
at the finite-temperature transition. On the other hand, simple
dimensional reduction suggests that the finite-temperature phase
transition is governed by order-parameter fluctuations in the $2d$
Ising universality class which is clearly different from mean-field
theory. This mean-field puzzle has first been studied in
\cite{Kogut:1998ri} using lattice simulations, followed by
  generalized large-$\Nf$ expansion techniques in
  \cite{Caracciolo:2005zu}. It turns out that its resolution requires
the existence of singularities in the critical exponents in the
large-$\Nf$ limit.

In the present work, we study the phase diagram of the
$(2+1)$-dimensional Gross-Neveu model as a function of the flavor
number and temperature. Because of asymptotic safety, this phase
diagram is a pure prediction of the theory once an overall scale
parameter is fixed. Our emphasis will be on universal aspects of
criticality, as well as near-critical aspects such as local-order
phenomena or the width or the critical region. Of particular interest
are the approach to the large-$\Nf$ limit as well as the expected
non-monotonic features at small flavor numbers. All these different
aspects can already be studied within a comparatively simple
nonperturbative approximation of the functional RG flow, whence we
concentrate in the present work on conceptual clarity rather than
quantitative precision. As our approximation scheme corresponds to
a next-lo-leading order result of a derivative expansion of the effective action, higher
quantitative precision can be expected from systematic inclusions of
higher orders. 

Since the physical spin of electrons in condensed-matter applications
of the Gross-Neveu model typically plays the role of flavor, the
$\Nf=2$ model is phenomenologically most important. For instance, the
excitonic \cite{Khveshchenko:2001zz} or antiferromagnetic instabilities
\cite{Herbut:2006cs} in graphite and graphene, respectively, can be 
associated with quantum phase transitions falling into this
universality class. Also the universal aspects of secondary $d$- to
$d+is$-wave pairing transitions in nodal $d$-wave superconductors are
expected to be determined by the $\Nf=2$ Gross-Neveu model. Therefore,
special $\Nf$-dependent structures of the many-flavor properties which
lie near $\Nf=2$ can also exert an influence on physical systems.

This work is organized as follows. In Sect.~\ref{sec:model}, we
introduce the model and discuss the relevant symmetries. Section
\ref{sec:meanfield} briefly summarizes the results of a standard
mean-field analysis, following earlier works in the literature, see,
e.g., Ref.~\cite{Rosenstein:1988pt}. Section \ref{sec:flow} is devoted to a
functional RG study of the model at next-to-leading order in a
derivative expansion. The many-flavor phase diagram at finite
temperature and further results are presented in Sect.~\ref{sec:results}. We
conclude in Sect.~\ref{sec:conc} and list a number of technically
relevant details needed for the RG analysis in the appendix.

%----------------------------------------------------------------
\section{Gross-Neveu model in $d=2+1$}
\label{sec:model}
%----------------------------------------------------------------

The classical action of the Gross-Neveu model in $d$ space-time dimensions reads
\be
\label{eq:fermionic_action}
S[\psibar,\psi]
&=&\int d^dx\left\{\sum_{j=1}^{\Nf}\barspinor\mathrm{i}\fslash{\partial}\,\spinor
  + \sum_{i,j=1}^{\Nf}\altbarspinor\altspinor\frac{\bar{g}}{2
    \Nf}\barspinor\spinor\right\}\nn\\ 
&\equiv& \int d^dx \left\{\psibar\mathrm{i}\fslash{\partial}\,\psi+ \frac{\bar{g}}{2 \Nf}(\psibar\psi)^2
\right\}\,,
\ee
where $\Nf$ determines the number of fermion flavors. The fermions interact via a 
four-fermion interaction term with the bare coupling given
by~$\bar{g}$ of mass dimension $[\bar{g}]=-1$. In our study of the Gross-Neveu model
in $d=2+1$, we shall use a reducible four-component ($d_{\gamma}=4$)
representation for the $\gamma$-matrices, as such Dirac spinors appear
naturally in the condensed-matter applications of this model, see
Refs.~\cite{Braun:2010tt,Scherer:2012PhD} for details of our conventions.

The symmetry which is most relevant for the present work is a
discrete $\mathds{Z}_2$ symmetry with generators given by
$\mathds{Z}_{2}^{5}=\{\mathds{1}_{4},\gamma_{5}\}$. The nontrivial
transformation of the Dirac spinors reads
\begin{equation}
\psi \to \gamma_5 \psi, \quad \bar\psi \to -\bar\psi
\gamma_5, \label{eq:trafo5}
\end{equation}
implying $\bar\psi\psi \to -\bar\psi\psi$ and thus leaving the action
\eqref{eq:fermionic_action} invariant. In agreement with the
literature, we will refer to this symmetry as a discrete ``chiral''
symmetry (even though it should rather be viewed as a subgroup of the
axial symmetry familiar from $4d$ models.) {Note that this} transformation acts
simultaneously on all flavors. 

In addition to this discrete chiral symmetry, the $(2+1)$-dimensional
Gross-Neveu model has a much larger continuous flavor symmetry: using
the product of Dirac matrices $\gamma_{35}:=i\gamma_3\gamma_5$, we can
define the projectors $P_{35,\pm}:=(1/2)(\mathds{1} \pm
\gamma_{35})$. For each of the corresponding spinor components
$\psi_{i,\pm}:=P_{35,\pm} \psi_i$, we can define a separate
U($\Nf$)$_\pm$ flavor rotation, such that the action
\eqref{eq:fermionic_action} is invariant under a
U($\Nf$)$_+\,\times\,$U($\Nf$)$_-$ continuous symmetry group. However,
the latter will not play any relevant role in the following and
remains unbroken in the discussed pattern of dynamical symmetry
breaking. {FRG studies of $3d$ chiral phase transitions with more
  complicated symmetry-breaking patterns can be found in
  \cite{Gies:2009da,Mesterhazy:2012ei,Gies:2010st}.} 

Perturbative arguments indicate that the Gross-Neveu field theory in $d=4$
dimensions is non-renormalizable. This implies that the theory has
to be defined with a physical cutoff, introducing -- strictly speaking
-- infinitely many physical parameters that specify the details of the
physical regularization. In lower dimensions $2<d<4$, the perturbative
reasoning is invalidated by the occurrence of a non-Gau\ss ian fixed
point in the RG flow that allows to define UV complete renormalized
trajectories that emanate from this interacting UV fixed point. This
{\it asymptotic safety} is conceptually reminiscent of UV completeness
due to asymptotic freedom with the only difference that the
(dimensionless) coupling approaches a finite fixed-point value in the
UV. A detailed analysis \cite{Braun:2010tt} shows, that a renormalized
trajectory is fully fixed by fixing an overall scale parameter. In
$2<d<4$, this can, for instance, be done by fixing the dimensionful
bare coupling $\bar{g}_\Lambda$ at an initial scale $\Lambda$. For
sufficiently strong initial coupling, the system is in the massive
chiral-symmetry broken (\xsb) phase such that the scale fixing can also be
done by choosing a value for the induced fermion mass $\mf$. 

In the present work, we will always assume that the initial coupling
has been chosen {super-critically, i.~e. the system} is in the
massive \xsb\ phase and all observables can be given in units of
$\mf$. {To compare different theories, we keep $\mf$ fixed to the 
same value for all $\Nf$.}
Incidentally, the critical coupling is directly connected with
the non-Gau\ss ian fixed-point coupling, implying a standard relation
between the RG fixed-point structure and critical aspects of the
chiral quantum phase transition. 

In passing, we note that the non-Gau\ss ian fixed point {of the four-fermion coupling} 
approaches the
non-interacting Gau\ss ian fixed point in the limit $d\to2$.
 In other words, asymptotic safety becomes asymptotic freedom in
$d=2$. Consequently, the phase transition disappears as the system
undergoes \xsb\ for any initial finite value of the coupling with mass
generation being associated with ``dimensional transmutation''. Even
though large parts of the analysis can be done in continuous
dimensions, we concentrate on $d=3$ in the following.

%----------------------------------------------------------------
\section{Mean-field analysis}
\label{sec:meanfield}
%----------------------------------------------------------------

Let us start by recapitulating a mean-field analysis of the model and
its finite-temperature phase transition. Such an analysis has already
been performed frequently in the literature, see,
e.g., Ref.~\cite{Rosenstein:1988pt}. We summarize here only the essentials
as a preparation for the functional RG anaysis. If only fermionic
fluctuations are considered as in the following, the mean-field
analysis is identical to the large-$\Nf$ limit. 

We begin by introducing an auxiliary bosonic Hubbard-Stratonovich
field $\sigma\sim \bar{\psi}\psi$, leading us to a so-called 
partially bosonized action:
\be S_{\rm PB}[\psibar,\psi,\sigma]=
\int_{x}\left\{\frac{\Nf}{2}\bar{m}^2\sigma^2
+\psibar\left(\mathrm{i}\fslash{\partial} 
    + \mathrm{i}\bar{h}\sigma\right)\psi\label{eq:PBaction} \right\}.
\ee
Since the $\sigma$ field occurs only quadratically, it can be integrated out exactly. This leads us back to the
fermionic action \eqref{eq:fermionic_action} if we identify
\begin{equation}
\bar{g} = \frac{\bar{h}^2}{\bar{m}^2}, \label{eq:bosferm}
\end{equation}
such that only the ratio $\frac{\bar{h}^2}{\bar{m}^2}$ has a physical
meaning at mean-field level. The $\sigma$ field in
Eq.~\eqref{eq:PBaction} transforms as $\sigma \mapsto -\sigma$ under
the discrete chiral transformation. Since its expectation value is
directly related to the fermion condensate~$\langle
\bar{\psi}\psi\rangle$, it serves as an order parameter for chiral
symmetry breaking and renders the fermions massive.  From a
study of the order parameter for a given~$\Nf$ we can obtain the 
critical temperature $\Tc$ and analyze the universal critical behavior
close to the phase {transition, see below.}

As the fermions occur only bilinearly in the partially bosonized
action \eqref{eq:PBaction}, they can be integrated out, yielding a
nonlocal bosonic effective action for the Gross-Neveu model,
\begin{equation}
S[\sigma]= \Nf \int_x \frac{1}{2} \bar{m}^2 \sigma^2 - \Nf \Tr \ln( i
\fslash{\partial} + i \bar{h} \sigma ). \label{eq:MFaction}
\end{equation}
The mean-field/large-$\Nf$ approximation is defined by ignoring fluctuations of the $\sigma$ field, i.e.,
\Eqref{eq:MFaction} defines the full theory at mean-field level. As we expect the ground state to be homogeneous for all
temperatures $T$ and flavor numbers $\Nf$, we restrict ourselves to
$\sigma=\sigma_0=\text{const.}$ here and in the
following.\footnote{Inhomogeneous ground states have been found in
the GN model in $d=2$ dimension~\cite{Schon:2000he}, but only in
the presence of a finite chemical potential. Indications for similar
phenomena in higher dimensions have also been {collected, see, e.g.,
\cite{Nickel:2009wj,Carignano:2012sx,Kojo:2011cn}.}} 

For constant $\sigma_0$, the effective action boils down to the
effective potential, $S[\sigma\to\sigma_0]= \int_x U(\sigma_0)$, with the ground
state {satisfying $\partial U(\sigma)/\partial \sigma |_{\sigma_0} =0$.} The 
solution is found from the variation of the action
\eqref{eq:MFaction}, yielding the gap equation,
\begin{equation}
\sigma_0=4 \frac{\bar{h}^2}{\bar{m}^2} \int \frac{d^3 p}{(2\pi)^3}
\frac{\sigma_0}{p^2 + \bar{h}^2 \sigma_0^2}.\label{eq:gap}
\end{equation}
Apart from the trivial solution $\sigma_0=0$, we find a nontrivial
solution being the global minimum of the effective potential for
sufficiently large initial coupling $\bar{g} = \bar{h}^2/\bar{m}^2$.
The precise evaluation of \Eqref{eq:gap} requires, of course, a
regularization of the linear UV divergence and a corresponding
renormalization of the parameters. Considering a sharp UV cutoff
$\Lambda\gg \bar{h} \sigma_0$ for simplicity, we adjust the initial
bare coupling $\bar{g} = \bar{h}^2/\bar{m}^2$ at the scale $\Lambda$
such that the induced fermion mass $\mfb= \bar{h} \sigma_0$ is kept
constant. Ignoring subleading orders in the cutoff, the
renormalization condition is given by
\begin{equation}
\mfb\simeq \frac{2}{\pi} \left( \Lambda - \frac{\pi^2}{2}
\frac{\bar{m}^2}{\bar{h}^2} \right) = \frac{2}{\pi} \Lambda  \left(1 - \frac{\pi^2}{2 \Lambda }
\frac{1}{\bar{g}} \right), \label{eq:MFmf}
\end{equation}
where we can read off the mean-field value of the (non-universal)
critical coupling for chiral symmetry breaking, $\bar{g}_{\text{cr}} =
\pi^2/(2\Lambda)$ (for a sharp cutoff). The effective potential can be
obtained by integrating the gap equation. For $\mfb\neq 0$, the result is, 
\begin{equation}
U(\sigma) = \int_0^\sigma d\sigma' \frac{\partial}{\partial \sigma'}
U(\sigma') = \frac{1}{\pi} \left( \frac{1}{3} \bar{h}^3 |\sigma|^3 -
\frac{\mfb}{2} \bar{h}^2 \sigma^2 \right), \label{eq:MFeffpot}
\end{equation}
where all the {scheme dependence is cancelled} by the corresponding
scheme dependence of the relation between the fermion mass $\mfb$ and
the initial couplings \cite{Cooper:2002yv}. Fixing the physical
observable $\mfb$, the regularization scheme defines a pair
$(\Lambda,\bar{g}(\Lambda))$, such that a different choice for the
cutoff $\Lambda$ is connected with a corresponding adjustment of
the initial coupling $\bar{g}(\Lambda)$, such that $\mfb$ is unaffected.

At finite temperature, the same analysis in the Matsubara formalism
requires to replace the momenta $p\to (\omega_n,\vec{p})$, with the
fermionic Matsubara frequencies $\omega_n=(2n+1)\pi T$, and the
spatial momenta $\vec{p}$. Performing the Matsubara sum, the
analogue of the gap equation \eqref{eq:gap} {(for $\sigma_0\neq 0$)}
reads \cite{Rosenstein:1988pt,Hands:1992ck}, 
\begin{equation}\label{eq:ge_finiteT}
1=4\,\bar{g} \int\!\! \frac{d^2 p}{(2\pi)^2}
\frac{1}{\sqrt{\vec{p}^{\,2}+\bar{h}^{2}\sigma^{2}}}
\left[1-2 n_{F}\left(\sqrt{\vec{p}^{\,2}+\bar{h}^{2}\sigma^{2}})\right)
\right]\!, 
\end{equation}
where $n_{F}(z)=1/(\mathrm{e}^{\beta z}+1)$ denotes the Fermi
distribution function at inverse temperature $\beta=1/T$. For the
system being in the chirally broken phase with fermion mass
$\mfb>0$ at zero temperature, the gap equation can be reexpressed 
in terms of the finite-temperature fermion mass,
\begin{equation}
\mfbT=\mfb -2 T \ln \left( 1+e^{\mfbT/T} \right). \label{eq:mfbT}
\end{equation}
For {fixed zero-temperature fermion mass $\mfb$, this transcendental equation can easily
be solved numerically for a given temperature, yielding a 
 thermal fermion mass $\mfbT$ which is
monotonically decreasing with temperature.} The mean-field critical temperature $\Tc$
where $\mfbT{}_{=\Tc}=0$ can analytically be read off from 
\Eqref{eq:mfbT}, 
\be 
\Tc=\frac{\mfb}{2 \ln 2}\simeq 0.7213\,\mfb. \label{eq:MFTc}
\ee
Analogously to \Eqref{eq:MFeffpot}, the thermodynamic effective
potential can be computed by integrating the gap equation in terms of
polylogarithms, yielding \cite{Cooper:2002yv,Caldas:2009zz}
\be
U(\sigma;T)\!\! &=&\!\!
\frac{1}{\pi}\left(\frac{1}{3}\bar{h}^{3}|\sigma|^{3}
-\frac{\mfb}{2}\bar{h}^{2}\sigma^{2}\right) \nonumber\\
&&\!\! +\frac{2\bar{h}|\sigma|T^2}{\pi} 
\mathrm{Li}_{2}\left(-\mathrm{e}^{-\bar{h}|\sigma|/T }\right)
 \label{eq:eff_pot}\\
&&\!\! +\frac{2T^3}{\pi}\mathrm{Li}_{3}\left(-\mathrm{e}^{-\bar{h}|\sigma|/T}\right)
+\frac{3T^3}{2 \pi}\zeta(3), \nonumber
\ee
where the last term involving the Riemann $\zeta$ function has been chosen
such that $U(\sigma=0;T)=0$. It is instructive to expand the effective
potential in powers of the mean field:
\be 
U(\sigma;T)  &=&   
\frac{\bar{h}^{2}\ln 2}{\pi}\left(T-\Tc\right)\sigma^{2} +
\frac{\bar{h}^{4}}{16\pi T}\sigma^{4}  \nonumber\\
&&- \frac{\bar{h}^{6}}{576\pi T^6}\sigma^{6} + \dots \,.
\label{eq:eff_pot_expansion}
\ee
Since only even powers of $\sigma$ appear, the effective potential is
analytic in the $\mathds{Z}_2$ invariant $\sigma^2$ at finite
temperature -- in contrast to the $T=0$ solution, where we observe a
$(\sigma^2)^{3/2}$ dependence, cf. \Eqref{eq:MFeffpot}. 

Finally, we can use these findings to extract thermodynamic critical
exponents in the mean-field approximation. The exponents $\alpha$ and
$\beta$ follow straightforwardly from the gap equation
\eqref{eq:ge_finiteT} and the effective potential \Eqref{eq:eff_pot},
\Eqref{eq:eff_pot_expansion}.  Expanding
\Eqref{eq:mfbT} close to $\Tc$ to
second order in $\mfbT$, we obtain the
relation \cite{Rosenstein:1988pt} 
\be 
\mfbT\simeq 2\sqrt{\mfb}\sqrt{\Tc-T},\label{eq:mbfTscal}
\ee
from which we can read off the scaling exponent of the order parameter
$\beta=1/2$. Plugging the result for the temperature-dependent ground
state into the effective potential, we obtain the free energy $f(T)$
per flavor species and per unit area. Expanding the thermal
contribution to fourth order in
$\mfbT$ and differentiating twice with respect to
temperature, the specific heat per unit area can be found
as \cite{Rosenstein:1988pt} 
\be c=-T\frac{\partial^{2}}{\partial T^{2}}f(T)\Big|_{T=\Tc}\simeq \frac{8
  (\ln 2)^{2}}{\pi}\Tc^{2}.  
\ee
Since the free energy vanishes in the symmetric phase, this implies
$\alpha=0$, i.e., a jump in the specific heat across the transition.
For the exponents $\delta$ and $\gamma$, the analysis has to be
repeated with the inclusion of an external source, whereas the
correlation exponent $\nu$ and the anomalous dimension $\eta_\sigma$
can be derived from the large-$\Nf$ limit of the induced bosonic
propagator. For instance, the inverse correlation length,
corresponding to the induced bosonic mass, $\xi^{-1}=m_\sigma$, turns
out to be proportional to the fermion mass, the temperature-scaling of
which has been worked out in \Eqref{eq:mbfTscal} above. The
correlation exponent $\xi \sim |\Tc-T|^{-\nu}$ thus yields the
standard mean-field value $\nu=1/2$. {As it should be, also} the other exponents yield
mean-field values, such that the critical behavior of the Gross-Neveu
model in $(2+1)$ dimensions in the large-$\Nf$ limit can be summarized by
\be\label{eq:mfexp}
\alpha=0,\,
\beta=\frac{1}{2},\,
\gamma=1,\,
\delta=3,\,
\nu=\frac{1}{2},\,
\eta_{\sigma}=0.
\ee
For further details, we refer to \cite{Rosenstein:1988pt}. These 
exponents do not satisfy hyperscaling relations, as can be expected
from the fact that bosonic fluctuations which establish such relations
are completely suppressed in the large-$\Nf$ limit. However, this
still raises a question concerning the analyticity of the large-$\Nf$
limit: at any finite flavor number, universality suggests that the
critical behavior of the thermal phase transition should be governed
by order-parameter fluctuations. Since the theory has a $\mathds{Z}_2$
symmetry, we expect the phase transition to be in the Ising
universality class. This is not reflected by the large-$\Nf$ limit
given by \Eqref{eq:mfexp}. Concerning the critical exponents 
for the transition to the high-temperature phase, the approach to the large-$\Nf$ limit must therefore
be non-analytic in $\Nf$.

Actually in all of the above-listed mean-field results, the dependence
{on the flavor number $\Nf$ naturally dropped out}. Non-trivial flavor
information thus requires the inclusion of both fermionic and bosonic fluctuations. 

%----------------------------------------------------------------
\section{Functional Renormalization Group Analysis}
\label{sec:flow}
%----------------------------------------------------------------

For our nonperturbative study of the many-flavor phase diagram of the
Gross-Neveu theory, we employ a functional RG equation for the quantum
effective action.  For this, we consider a scale-dependent effective
action $\Gamma_k$, governing the field dynamics at an infrared (IR)
scale~$k$. The so-called effective average action $\Gamma_k$ already
comprises all quantum fluctuations from the UV cutoff $\Lambda$ down
to $k$. In the limit $k\to 0$, the effective average action is
identical to the full quantum effective action, i.e., the 1PI
generating functional, whereas it approaches the classical action in
the limit $k\to \Lambda$. The $k$ evolution of $\Gamma_k$ is
determined by the Wetterich equation~\cite{Wetterich:1992yh},
\begin{equation}\label{eq:flowequation}
	\partial_t\Gamma_k[\Phi]
        =\frac{1}{2}\mathrm{STr}\Bigl\{\bigl[\Gamma^{(2)}_k[\Phi]+R_k\bigr]^{-1}(\partial_tR_k)\Bigr\},
        \;\, \pat=k\frac{d}{dk}.
\end{equation}
The quantity $\Gamma^{(2)}_k$ denotes the second functional derivative
with respect to the field variable $\Phi$, collecting all fermionic and bosonic fields. The
momentum-dependent function $R_k$ is a regulator, suppressing IR modes
below a scale
$k$. For reviews of the functional RG adapted to the present context, see 
Refs.~\cite{Berges:2000ew,Aoki:2000wm,Pawlowski:2005xe,Gies:2006wv,
kopietz2010introduction,Metzner:2011cw,Braun:2011pp}. 
At finite temperature, the effective action
is directly related to thermodynamic quantities. For instance, the
corresponding effective potential at its minimum corresponds to
minus the pressure. 
{In the present work, we intend to study the many-flavor phase diagram
in a nonperturbative fashion using our continuum RG approach}. 
For this, we use as an ansatz for the effective action:
\be
\Gamma_k[\psibar,\psi,\sigma]&=&\int_{x}\Bigl\{\frac{{\Nf}}{2}Z_{\sigma}(\partial_{\mu}\sigma)^2+
\psibar\left(Z_{\psi}\mathrm{i}\fslash{\partial} + \mathrm{i}\bar{h}\sigma\right)\psi\nonumber\\
&&\qquad + \Nf\, U(\sigma^2)\Bigr\}\,,
\label{eq:PBgamma}
\ee
where~$U$ denotes the effective potential for the order parameter. The
couplings and the wave-function renormalizations~$Z_{\sigma}$
and~$Z_{\psi}$ are assumed to be scale dependent. This ansatz is
related to a systematic derivative expansion at next-to-leading order
and has already proved to quantitatively describe the quantum-phase
transition at zero temperature
\cite{Rosa:2000ju,Hofling:2002hj,Braun:2010tt}. For simplicity, the
present ansatz ignores possible anisotropic wave function
renormalizations due to the presence of the heat bath, see, e.g.,
\cite{Braun:2009si}.

At the cutoff scale~$k=\Lambda$, the initial conditions for the
couplings are fixed such that the action~\eqref{eq:fermionic_action}
is recovered at this scale. This is ensured by the choice
\begin{equation}
Z_{\sigma,\Lambda} \ll 1, \,\, Z_{\psi,\Lambda} = 1,\,\, 
\bar{h}_\Lambda^2=\bar{g}_\Lambda \bar{m}_\Lambda^2,\,\,
U_{\Lambda}(\sigma^2)=\frac{1}{2}\bar{m}^{2}_{\Lambda} \sigma^2.
\end{equation}

\begin{figure}[t]
\begin{center}
\includegraphics[width=1.0\linewidth]{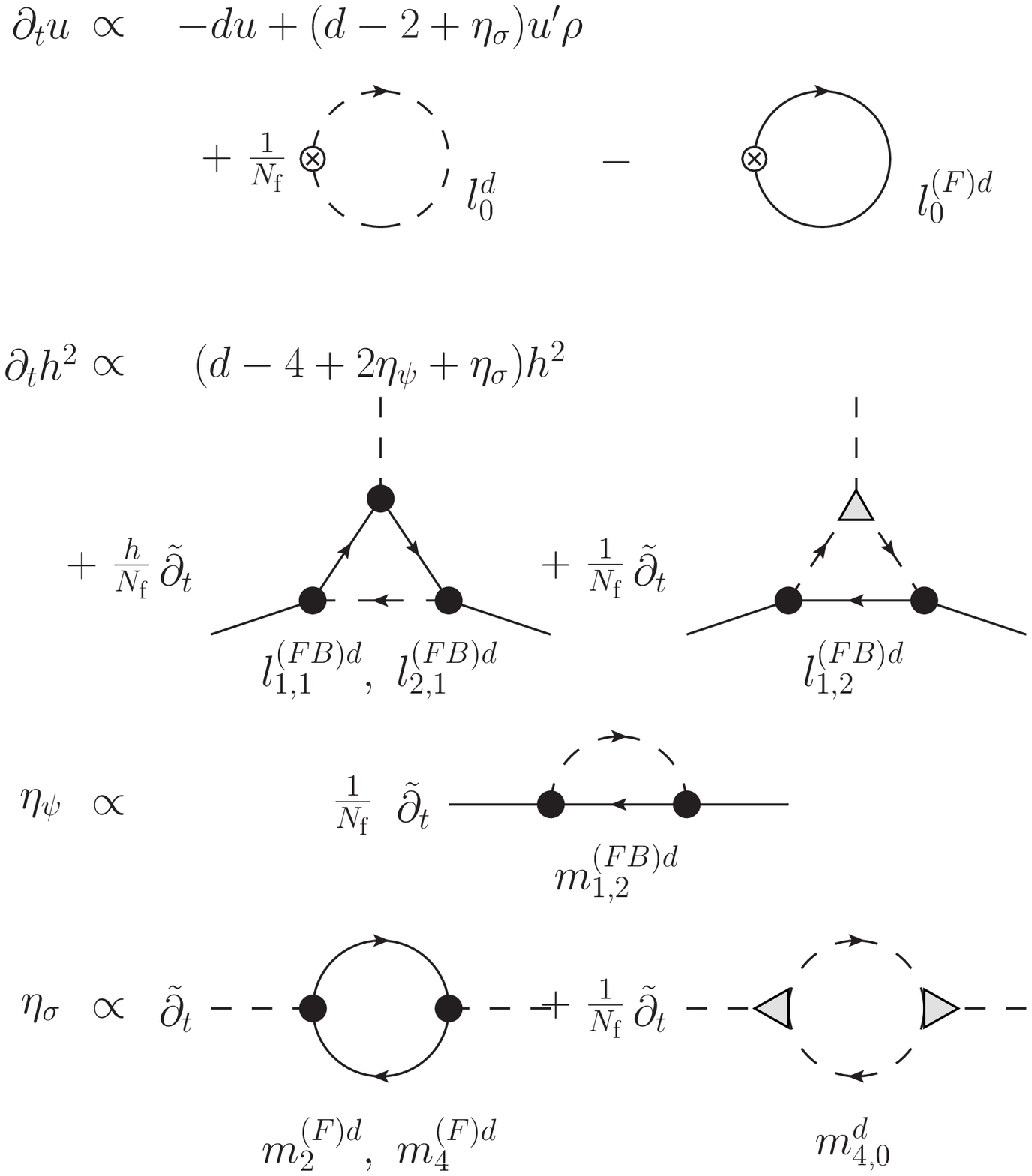}
\end{center}
\caption{Diagrammatic representation of the flow equations \Eqref{eq:u_flow}, \Eqref{eq:hflow}, \Eqref{eq:etasigma}
and \Eqref{eq:etapsi}. The {1PI} diagrams represent quantum fluctuations renormalizing
the vertex functions at the RG scale $k$. Full (dashed) lines represent our
ansatz for the fermionic (bosonic) propagator modified by a regulator term $R_{k}$.
Black disks denote the Yukawa vertex, while gray triangles denote vertices $\sim U^{\prime\prime\prime}(\sigma)$
which contribute only in the symmetry-broken regime of the flow.
Below the {1PI diagrams, the labels ($l$, $m$)  of the associated threshold functions are given, see appendix for details.} 
The $\tilde{\partial_{t}}$-derivative
acts on the regulator dependence of the regularized propagators and yields regulator insertions $\partial_{k}R_{k}$, as denoted
by crossed circles in the flow-diagrams for the effective potential. Note that all diagrams
with an internal bosonic line are suppressed by factors of $1/\Nf$.}
\label{fig:diagrams}
\end{figure}

From our ansatz~\eqref{eq:PBgamma}, we obtain the following flow
equation for the dimensionless order-parameter
potential~$u(\rho)=k^{-d}U(\sigma^2)$:
\be
\partial_{t}u  &=&   -d u + (d-2+\eta_{\sigma})u^{\prime}\rho -2d_{\gamma}v_d\, l_0^{{(F)}d}(2 h^2\rho;\eta_{\psi}) \nn\\ 
&& \qquad+ \frac{1}{\Nf} 2 v_d \,l_0^{d}(u^{\prime}+2\,\rho
u^{\prime\prime};\eta_{\sigma})\,, 
\label{eq:patu}
\ee
where 
\be
\rho=\frac{1}{2}Z_{\sigma}k^{2-d}\sigma^{2},\label{eq:u_flow}
\ee
and the dimensionless renormalized Yukawa coupling 
\be
 h^{2}=Z_{\sigma}^{-1}\wavpsi^{-2} k^{d-4}\bar{h}^{2}.
\ee
The corresponding flow for the Yukawa coupling reads,
\begin{widetext}
\be\label{eq:hflow}
\partial_{t}h^2 & = &  (d-4 + 2\eta_{\psi}+\eta_{\sigma})h^2 +% \nn \\
%& & 
 \frac{1}{N_{\mathrm{f}}}8 v_d\,h^4 \, l_{1,1}^{(FB)d}(2
 h^2\kappa,\left[u^{\prime}(\kappa)+2\kappa
   u^{\prime\prime}(\kappa)\right];\eta_{\psi},\eta_{\sigma})
  \nn \\ 
& & - \frac{1}{N_{\mathrm{f}}}\left(48\kappa
  u^{\prime\prime}(\kappa)+32\kappa^{2}u^{\prime\prime\prime}(\kappa)\right)
  v_{d}\,h^{4}\,l_{1,2}^{(FB)d}(2
  h^2\kappa,\left[u^{\prime}(\kappa)+2\kappa
    u^{\prime\prime}(\kappa)\right];\eta_{\psi},\eta_{\sigma})
   \nn \\ 
& & - \frac{1}{N_{\mathrm{f}}}32 v_{d}\,
   h^{6}\,\kappa\,l_{2,1}^{(FB)d}(2
   h^2\kappa,\left[u^{\prime}(\kappa)+2\kappa
     u^{\prime\prime}(\kappa)\right];\eta_{\psi},\eta_{\sigma}),
\ee
where the right-hand side is understood to be evaluated at the
$k$-dependent minimum of the effective potential, i.e., $\rho_{\rm min}=\kappa=0$ in
the symmetric and $\rho_{\rm min}=\kappa\neq 0$ in the symmetry-broken regime. Here,
$\kappa$ is related to the dimensionless renormalized expectation
value of the scalar field, 
\be
\kappa=\frac{1}{2}\frac{Z_{\sigma}\sigma_0^2}{k^{d-2}}\,,
\ee
satisfying $u'(\kappa)=0$ in the broken
regime. The same conventions apply to the anomalous
dimensions
\be
\eta_{\psi} & = & \frac{1}{N_{\mathrm{f}}} 8 \frac{v_{d}}{d}h^{2}\,m_{1,2}^{(FB)d}(2 h^2\kappa,\left[u^{\prime}(\kappa)+2\kappa u^{\prime\prime}(\kappa)\right];\eta_{\psi},\eta_{\sigma})\,,\label{eq:etapsi}\\
\eta_{\sigma} & = &  8
\frac{d_{\gamma}v_d}{d}h^2\left[m_{4}^{(F)d}(2
  h^2\kappa;\eta_{\psi})-2h^{2}\kappa m_{2}^{(F)d}(2 h^2\kappa;\eta_{\psi})\right] + \nn \\
& &  \frac{1}{N_{\mathrm{f}}}8\frac{v_{d}}{d}\kappa \left(3u^{\prime\prime}(\kappa)+2\kappa u^{\prime\prime\prime}(\kappa)\right)^{2}m_{4,0}^{d}(\left[u^{\prime}(\kappa)+2\kappa u^{\prime\prime}(\kappa)\right],0;\eta_{\sigma})\,.\label{eq:etasigma}
\ee
\end{widetext}
All so-called threshold functions $l$, $m$ occurring in these flow equations depend
on the details of the regulator. Formulas for general regulators as
well as explicit expressions for the linear regulator
\cite{Litim:2000ci}  used in this work are
given in the appendix. These threshold functions describe the behavior
of the regularized 1PI diagrams across mass thresholds and thus
describe the decoupling of massive modes from the flow. They also
carry the temperature dependence and thus encode also the decoupling
of higher Matsubara frequencies as a function of the dimensionless
temperature $\tau=2\pi T/ k$. A diagrammatic representation of the flow 
equations corresponding to our ansatz for the quantum effective action is
given in Fig.~\ref{fig:diagrams}.
%
%--------------------------Figure--------------------------(fRG diagrams)
%----------------------------------------------------------
%

%----------------------------------------------------------------
\section{Many-Flavor Phase Diagram and Critical Behavior}
\label{sec:results}
%----------------------------------------------------------------
 
In order to keep the present analysis simple, we solve the flow of the
potential in a straightforward polynomial expansion,
\begin{equation}
u(\rho)=\sum_{{n=1}}^{\infty}\frac{\lambda_{2n}}{n!}(\rho)^n\quad\text{and}\,\,
u(\rho)=\sum_{n=2}^{\infty}\frac{\lambda_{2n}}{n!}(\rho-\kappa)^n.
\end{equation}
in the symmetric regime and broken regime ($\kappa >0$), respectively. 
Moreover, we ignore higher order
couplings with $n\geq3$. In fact, higher-order approximations, though
widely used in the literature, appear to induce artificial
instabilities in the present model in a (physically less relevant)
temperature interval below $\Tc(\Nf)$. We have checked that this
approximation does not oversimplify the model by verifying all known
limits. For instance, aspects of the large-$\Nf$/mean-field limit such
as the critical temperature are reproduced by our simple approximation
with an error of $\sim 3$\% ($\Tc\simeq 0.7439\,\mfb$ in comparison with
the exact result $\Tc\simeq 0.7213\,\mfb$, cf. \Eqref{eq:MFTc}). We 
assume that this systematic error is also indicative for the error
size at finite $\Nf$.

The following results are obtained from numerically integrating the
RG flow in the plane of temperatures and flavor numbers $(T,\Nf)$. Apart
from the errors of the numerical routines which are much smaller than
the above-mentioned systematic error, we also have to make sure that
the initial conditions of the flow lie on the ``renormalized
trajectory'', i.e., the line of constant physics
$(\Lambda,g(\Lambda)$, that relates initial conditions for different
choices of the cutoff $\Lambda$ to one and the same
physical system in the long-range limit. Moreover, this trajectory
must also be on top of or at least sufficiently close to the critical
hypersurface emanating from the non-Gau\ss ian fixed point in order to
interconnect the long-range limit with the UV complete
theory.\footnote{Whereas chosing a trajectory exactly on the critical
  hypersurface is essential for a UV-complete limit, it is numerically
  not relevant for the long-range physics, as trajectories in the
  attractive domain of the fixed point are attracted towards the
  critical hypersurface exponentially fast.} 
%
%--------------------------Table--------------------------(compare Tc to literature)
\begin{table*}[t]\center
\begin{tabular}{p{65pt}||p{65pt}|p{65pt}||p{65pt}|p{65pt}|p{65pt}p{0pt}}
\centering  & \centering $\Nf=4$ \\ (MC)~\cite{Kogut:1999um} &
\centering $\Nf=4$ \\ (FRG) & \centering $\Nf=12$
\\ (MC)~\cite{Hands:1998jg} & \centering $\Nf=12$ \\ (PTRG)~\cite{Castorina:2003kq} & \centering $\Nf=12$\\ (FRG) & \\ \hline\hline
\centering  $\frac{\Tc(\Nf)}{\mf}$ & \centering 0.49  &
\centering 0.52 & \centering 0.66$\pm$ 0.05 & \centering 0.57 &  \centering 0.62 & 
\end{tabular}
\caption{Comparison of our
  results (FRG) for $\Tc(\Nf)$ to those of proper-time renormalization
  group flows (PTRG)~\cite{Castorina:2003kq} and Monte Carlo
  simulations (MC)~\cite{Hands:1998jg,Kogut:1999um}. The
    conversion of the lattice data of \cite{Hands:1998jg} for
    $\Tc/\sigma_0$ for $\Nf=12$ to our observable $\Tc/\mf$ assumes
    that a potential multiplicative renormalization constant is close
    to unity.}
\label{tab:tc}
\end{table*}
%---------------------------------------------------------
%
In practice, we keep the UV cutoff fixed at
$\Lambda=100\,$a.u. ({auxiliary} units). Then we choose two different
sets of initial parameters for the Yukawa coupling and the {mass parameter} that
are close to the critical manifold. Integrating the flow down towards
the IR and expressing all dimensionful quantities in terms of the same
scale, such as the renormalized fermion mass,
\begin{equation}
\mf=\frac{1}{Z_\psi} \bar{h} \sigma_0\big|_{k\to 0} = \sqrt{2 k^2 h^2
  \kappa}\big|_{k\to 0}, \label{eq:mfrenorm}
\end{equation}
has to lead to quantitatively identical results.\footnote{Note that at
the mean-field level $Z_{\psi}=1$, rendering the renormalization factor trivial.
In that case the notions of renormalized and unrenormalized mass coincide.}
The first set (set I) of
parameters consists of choosing $h_\Lambda = 18.0$ and tuning the bare
bosonic mass term $\lambda_{2,\Lambda}$ such that $\mf=1\,$a.u. In the
second set (set II), we choose $h_\Lambda=6.0$ and tune the bare boson mass
such that $\mf=5\,$a.u. The initial bosonic wave function
renormalization is set to $Z_{\sigma,\Lambda}=1.0\times 10^{-3}$,
{such that the bosonic field is (almost) purely auxiliary at
  $\Lambda$ and the corresponding four-fermion vertex is (almost) pointlike}. Any difference between
the results for these two different sets are due to nonuniversal
finite cutoff effects, the influence of irrelevant operators and the
corresponding limitations of the truncation. Again, the differences
which we find are smaller than the above-mentioned systematic errors.

%
%--------------------------Figure--------------------------(phase diagram/fit to phase boundary)
\begin{figure}[t]
\begin{center}
\includegraphics[width=1\linewidth]{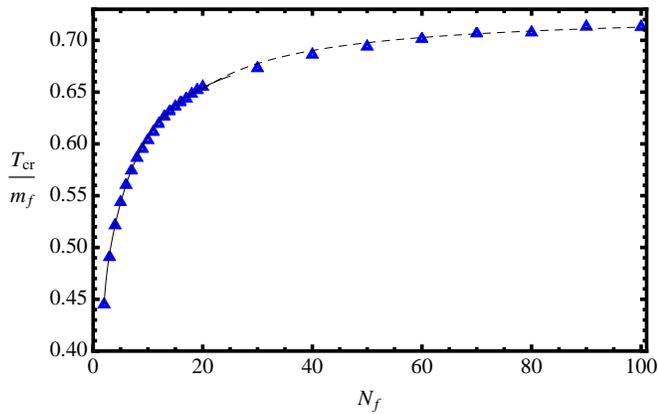}
\end{center}
\caption{Phase boundary of the Gross-Neveu many-flavor phase diagram as within our ansatz as obtained from set I of initial conditions. We performed temperature scans for flavors $N_{\mathrm{f}}=1,2,3,\dots,19$ and $N_{\mathrm{f}}=20,30,\dots,100$. Also shown are fits to the phase boundary in the large-$\Nf$ and small-$\Nf$ region, see~\Eqref{eq:fitTA} and~\Eqref{eq:fitTB}.}
\label{fig:PD1}
\end{figure}
%----------------------------------------------------------
%
A typical flow at $T=0$ starts at $k=\Lambda$ in the symmetric regime, and is
then driven by fermion fluctuations into the broken regime where the
flow freezes out, because of mass generation. At finite $\Nf$, the
fermion fluctuations are somewhat counterbalanced by the boson
fluctuations.

{Loosely speaking, finite temperature $T>0$ favors bosonic fluctuations. This is
a generic mechanism as the fermions acquire non-zero 
thermal masses because of the anti-periodic boundary conditions,
whereas this thermal decoupling does not apply to the zeroth Matsubara
mode of the bosons. For $0<T<\Tc$, the zero-temperature condensate
generated by the fermions becomes somewhat reduced by the bosonic
zeroth Matsubara mode. In this picture, $\Tc$ corresponds to the
temperature where the bosonic zeroth Matsubara mode is capable of
driving the scalar condensate to zero again such that the flow ends
again in the symmetric regime. From this point of view, it is also
clear that the flow even enters the broken regime at intermediate
scales for $T\gtrsim\Tc$, but, of course, ends up in the
symmetric regime. This phenomenon of local order is also generic
for Yukawa-like systems both in relativistic \cite{Braun:2009si} as well as in
non-relativistic systems \cite{Diehl:2007th}. 

Our results for the phase diagram in the $(T,\Nf)$ plane up to
$\Nf=100$ are shown in Fig.~\ref{fig:PD1}.  Also shown is a fit to the
large-$\Nf$ behavior (dashed line)
\begin{equation}
\frac{\Tc(\Nf)}{\mf}\simeq 0.72686 - 1.55098 \frac{1}{\Nf}, \quad
(\text{fit:}\, \Nf=20,\dots,100). \label{eq:fitTA} 
\end{equation}
Here, we did not fix the leading order term ($\Nf^0$) to the
  exact mean-field value, see Eq.~\eqref{eq:MFTc}. The next-to-leading
  order (NLO) term $\sim 1/\Nf$ is in principle computable with
  large-$\Nf$ techniques \cite{Caracciolo:2005zu}, though no explicit
  value is known. The NNLO term has been determined to scale as $(\ln
  \Nf)/ \Nf^2$, see Ref.~\cite{Caracciolo:2005zu}. 

For practical purposes, we also present a fit to the smaller $\Nf$ region (solid line),
\be
\frac{\Tc(\Nf)}{\mf}& \simeq & 0.695994 - \frac{1.11289}{\Nf} +
\frac{2.18236}{\Nf^2}
 - \frac{2.43492}{\Nf^3}\nonumber\\
&& + \frac{1.04127}{\Nf^4},  \quad (\text{fit:}\,
 \Nf=2,\dots,19).\label{eq:fitTB} 
\ee
In Fig.~\ref{fig:PDcomp} as well as in Tab.~\ref{tab:tc}, we perform
a comparison to existing results in the literature.
%
%--------------------------Figure--------------------------(phase diagram comparison)
\begin{figure}[t]
\begin{center}
\includegraphics[width=1\linewidth]{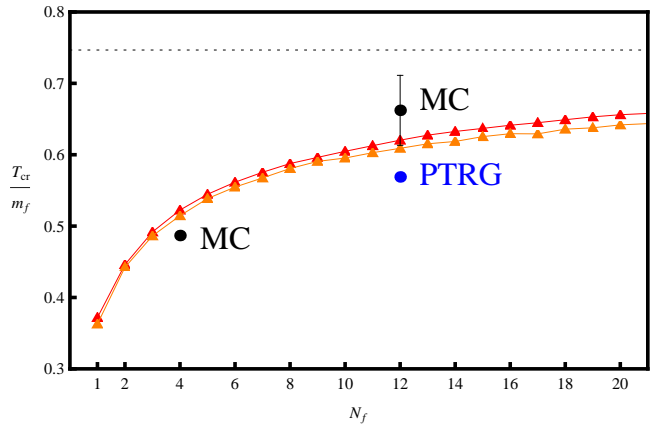}
\end{center}
\caption{The phase boundary obtained from our flow
  equations (red triangles corresponding to set I and orange triangles
  corresponding to set II, respectively) in comparison to available
  results from proper-time renormalization group (PTRG,
  $\Nf=12$)~\cite{Castorina:2003kq} and Monte Carlo simulations (MC,
  $\Nf=4,\,12$)~\cite{Hands:1998jg,Kogut:1999um}.}
\label{fig:PDcomp}
\end{figure}
%----------------------------------------------------------
%
We compare our FRG results to those from proper-time flows
(PTRG)~\cite{Castorina:2003kq} and predictions from two different
Monte Carlo simulations (MC)~\cite{Hands:1998jg,Kogut:1999um}. 
{While \cite{Hands:1998jg} provides for a statistical error,} 
an estimate of the systematic errors of each calculation is not
available. For
instance, the PTRG results are on the one hand obtained from a
larger truncation than the present one, but on the other hand do not
rely on an exact RG flow~\cite{Litim:2002xm}. {The $\Nf=4$ calculation
  of \cite{Kogut:1999um} provides data for $\Tc/\mf$ but no
  concrete error information. The $\Nf=4$ calculation
  \cite{Hands:1998jg} provides data with a statistical error for
  $\Tc/\sigma_0$ which can correspond to $\Tc/\mf$ within the
  conventions of \cite{Hands:1998jg} provided a multiplicative
  renormalization is close to unity. }
Our results differ from MC \cite{Kogut:1999um} and
PTRG results by $6\%-8\%$, which is a deviation slightly larger than
our expected systematic error, {whereas we find consistency with
  the MC data of \cite{Hands:1998jg} within error bars}. On the other
hand, our curve appears to be a compromise between the existing
results so far, {see Fig.~\ref{fig:PDcomp}}.  
%
%--------------------------Figure--------------------------(local order)
\begin{figure}[t]
\begin{center}
\includegraphics[width=0.96\linewidth]{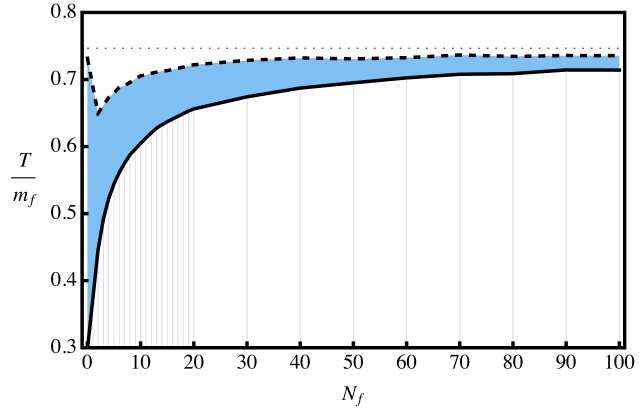}
\end{center}
\caption{Local order in the Gross-Neveu model: the shaded local-order
  region is bounded by the critical temperature $\Tc(\Nf)$ (solid
  line) and the temperature $T^\ast(\Nf)$ (dashed line) above which
  the system is in the disordered regime on all scales. In the
  local-order region, a non-vanishing condensate is found at
  intermediate $k$ scales, which is ultimately driven to zero by the
  long-range bosonic fluctuations.  Interestingly, the $\Nf= 1$ Gross-Neveu
  model shows an enlarged temperature window compared to theories with
  $\Nf >1$.  }
\label{fig:localorder}
\end{figure}
%----------------------------------------------------------
%

Let us now quantify the phenomon of local order introduced in the preceding discussion. As
mentioned already, even above the critical temperature there is a
temperature regime $\Tc(\Nf)<T<T^\ast(\Nf)$, where the flow first
enters the broken regime at intermediate scales $k$ but then re-enters
the symmetric regime in the limit $k\to 0$. This implies that the flow
exhibits a (nonuniversal) finite condensate at these intermediate
scales which is eventually depleted by the long-range bosonic
fluctuations. Identifying $1/k$ with a typical length scale, this
intermediate condensate can be interpreted as the existence of local
or short-range order on these typical length scales in the
system.\footnote{For example, the inverse IR regulator scale~$1/k$ can be roughly associated with the side length~$L$ 
of the spatial volume in lattice simulations~\cite{Braun:2004yk}.}  
 Only for higher temperatures $T>T^\ast(\Nf)$, no such local
order is observed and the system is in the disordered state on all
scales. Our results for $T^\ast(\Nf)$ from numerically solving the
flow are depicted in Fig.~\ref{fig:localorder} (dashed line) in comparison
with the critical temperature $\Tc(\Nf)$ (solid line).  It is
interesting to observe that the $\Nf= 1$ Gross-Neveu model shows an
enlarged temperature window for local order compared to theories with
$\Nf >1$, see Fig.~\ref{fig:localorder}. This might be due to the fact
that already the UV fixed point potential is closer to the broken
regime. The temperature $T^\ast(\Nf)$ lies always below the mean field
critical temperature (dotted line). This is only natural, as
the local-order phenomenon arises from an interplay of fermionic and
bosonic fluctuations whereas the mean-field approximation ignores
bosonic fluctuations {completely, see our discussion above.}

{At this point, we would like to add that a RG study of the Gross-Neveu model in a widely-used local
potential approximation (i.~e. setting $Z_{\sigma}\equiv 1$ and~$Z_{\psi}\equiv 1$) is not meaningful since it is spoilt 
by a dramatic parameter dependence. To be more specific, the parameters can be adjusted such that we obtain the same
fermion mass in the vacuum for all~$\Nf$ but the well-known exact result for the critical temperature in the large-$\Nf$ limit
is nevertheless not reproduced correctly, i.~e. it differs, e.~g., by a factor of two from the well-known result. This observation 
can be traced back to the fact that the fixed-point structure of the theory is oversimplified in this approximation.
}

We now turn to potential non-analyticities of the large-$\Nf$ limit.
On the one hand, the large-$\Nf$ limit discussed in
Sect.~\ref{sec:meanfield} {\it per constructionem} predicts that the finite-temperature
phase transition exhibits a mean-field behavior with mean-field
critical exponents. On the other hand, general universality arguments
suggest that at any finite $\Nf$, the phase transition should be in
the $2d$-Ising universality class, since the scalar order parameter measures
the breaking of a discrete $\mathds{Z}_2$ symmetry. Also, close to the
phase transition, the standard dimensional-reduction mechanism of the
Matsubara formalism should be at work, distinguishing the bosonic
zero-mode as the driving mode for the approach to criticality. 

The RG flow has to be able to resolve the obvious tension between
these two apparently contradictory viewpoints. One possibility is that
this contradiction is resolved by a non-analytic dependence of the
critical phenomena on $\Nf$ potentially arising from a failure
  of the suppression of IR fluctuations with increasing $\Nf$
  \cite{Chandrasekharan:2004uw}. The other, less dramatic possibility
is that the size of the critical region close to the phase transition
scales (possibly non-analytically) with increasing $\Nf$ to zero,
leaving us with the mean-field predictions in the $\Nf\to \infty$
limit. Evidence for the latter possibility has been found in
\cite{Kogut:1998ri} and \cite{Caracciolo:2005zu}. In
\cite{Kogut:1998ri}, $2d$-Ising scaling has been verified by measuring
the critical exponents very close to the phase transition with lattice
simulations for $\Nf=4$, $12$, $24$. On the other hand, beyond this
regime a region obeying mean-field scaling has been identified and the
location of the break-down of mean-field scaling in terms of a value
for the condensate $\sigma_{\text{MF}}$ has been determined. More
precisely, for temperatures $T<\Tc$ corresponding to a condensate near
$\sigma_0\gtrsim\sigma_{0,\text{MF}}$, mean-field scaling is observed,
whereas strong deviations from mean-field scaling and a crossover
towards the $2d$-Ising regime occurs for $\sigma<\sigma_{\text{MF}}$,
corresponding to temperatures closer to $\Tc$. In
Ref.~\cite{Kogut:1998ri}, the value of $\sigma_{\text{MF}}$ has been
shown to scale as $\sigma_{\text{MF}}\sim 1/\Nf^{x_{\text{MF}}}$. The
numerical value for the exponent $x_{\text{MF}}\simeq0.51$ has been
measured, nicely corresponding to an analytical estimate
$x_{\text{MF}}=1/2$ which was derived from a simple estimate for the
validity region of the mean-field approximation \cite{Kogut:1998ri}.

In \cite{Caracciolo:2005zu}, generalized large-$\Nf$ techniques
  have been used to compute the scaling of a fiducial relative
  temperature $T-\Tc(\Nf)$ inside the Ginzburg critical region as a
  function of $\Nf$, yielding $T-\Tc(\Nf) \sim 1/\Nf^{x_\text{G}}$
  with $x_{\text{G}}=1$. It was further argued that a scaling of the
  mean-field region measured in terms of $\sigma_{\text{MF}}$ with
  $x_{\text{MF}}=1/2$ and of the Ginzburg region measured in terms of
  the temperature with $x_{\text{G}}=1$ is compatible with each other
  within large-$\Nf$ considerations.

In the present work, we go one step further and compute the
  condensate for a wider range of temperatures, resolving both the
  mean-field as well as the Ginzburg region. In particular, we
directly determine the size of the Ginzburg region around the phase
transition. We do so by quantifying the regime where $2d$-Ising
scaling is observed. For consistency reasons, the Ginzburg region must
scale to zero with $\Nf\to \infty$ as fast as or even faster than the
validity region of the mean-field approximation. The size and scaling
of the Ginzburg region is an important piece of information for
{experimental} searches for the true critical region of the 2nd order
phase transition.
%
%--------------------------Figure--------------------------(local order)
\begin{figure}[t]
\begin{center}
\includegraphics[width=0.96\linewidth]{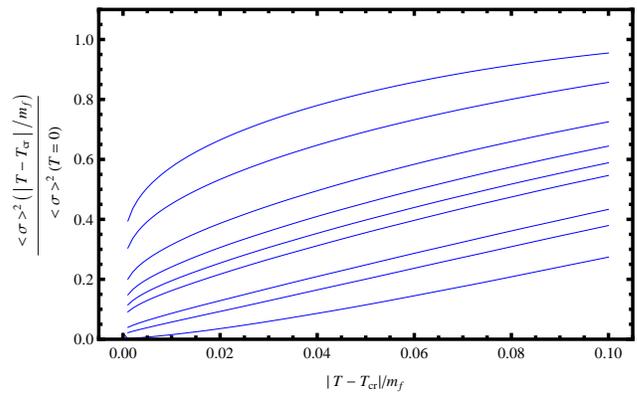}
\end{center}
\caption{The normalized order parameter squared as a function of 
temperature close to the phase transition. From top to bottom, the 
curves correspond to $\Nf=1,\,2,\,4,\,6,\,8,\,10,\,20,\,30\,,100$. For small
$\Nf$, the data is not well described by {mean-field theory
  corresponding to a linear function in this plot.}
However, as the flavor number increases,
the temperature dependence is predominantly linear, as expected from the 
mean-field scaling with $\beta_{\rm MF}=1/2$. Deviations to linear scaling
occur only upon approaching the critical temperature. As in \cite{Kogut:1998ri}, this
suggests an increase of the width of the mean-field scaling region with increasing
flavor number.}
\label{fig:sigmaMF}
\end{figure}
%----------------------------------------------------------
%
In order to identify the Ginzburg region, we use the following
strategy: The chiral condensate near the second-order phase
transition scales as
\begin{equation}
\langle \sigma \rangle \sim |T-\Tc|^\beta, \label{eq:GinzburgID}
\end{equation}
where $\beta$ in the exact solution should correspond to the exact
$2d$-Ising critical exponent $\beta=1/8$. The present FRG study indeed
boils down to that of a $2d$ scalar $\mathds{Z}_2$-symmetric model
near $\Tc$ due to dimensional reduction, guaranteeing that we are
analyzing the correct universality class. However, our simple
truncation is not capable of reproducing the exact Onsager critical
exponents. Therefore, we identify the Ginzburg region by identifying
the region in the phase diagram where \Eqref{eq:GinzburgID} holds with
a value for $\beta$ which matches that {of} a $2d$
$\mathds{Z}_2$ theory in a $\sigma^4$-truncation at NLO in a
derivative expansion (with a linear regulator). For this, we find
$\beta_{\text{trunc}}=0.0931$ using completely standard functional FRG techniques
\cite{Berges:2000ew}. At fixed $\Nf$, a one parameter fit of the form
$\langle \sigma \rangle =A |T-\Tc|^{\beta_{\text{trunc}}}$ with fit
parameter $A$ is then applied to a pre-selected region very close to
the critical temperature, where the power-law is satisfied by the
FRG data. For some flavor numbers, the FRG data is not in perfect
agreement with the power-law, possibly due to inaccuracy in the
determination of the critical temperature. {Our predicted 
values for the size of the 
Ginzburg region should thus be taken with some care.} 

Next, we define the temperature $T_{\text{G}}(\Nf)$
characterizing the size of the Ginzburg region by requiring the
relative deviation of the one-parameter {fit to the} FRG data to be
less than a few percent inside that region. More precisely, we performed
the determination of $T_{\text{G}}(\Nf)$ from our data with several criteria
corresponding to $2\%$, $3\%$, $4\%$, $5\%$ and finally $6\%$ tolerance. 

Repeating the analysis for various
$\Nf$, we find that we can fit the corresponding width $w_\beta(\Nf)$ of
the Ginzburg region scales to a powerlaw
\begin{equation}
w_\beta (\Nf)= \left| \frac{T_{\text{G}}(\Nf) - \Tc(\Nf)}{\mf}
\right| \sim \frac{1}{\Nf^{x_{\text{G}}}},\label{eq:wbeta}
\end{equation}
where the exponent obtained from the fit varies from
$x_{\text{G}}\simeq 0.93$ to $x_{\text{G}}\simeq 1.06$ in the range of
tolerances from $2\%$ to $6\%$. This apparent definition
  dependence of $x_{\text{G}}$ may also originate from the influence
  of subleading powers of $1/\Nf$ which need not be vanishingly small
  in the fit regime. Still, our result is well compatible with the
  generalized large-$\Nf$ result $x_{\text{G}}=1$
  \cite{Caracciolo:2005zu}. In this sense, it is remarkable that this
  scaling law appears to hold down to rather small values of $\Nf$}. A
plot of the scaling laws is shown in Fig.~\ref{fig:Ginzburg}.
%
%--------------------------Figure--------------------------(log-log-plot Ginzburg region)
\begin{figure}[t]
\begin{center}
\includegraphics[width=1\linewidth]{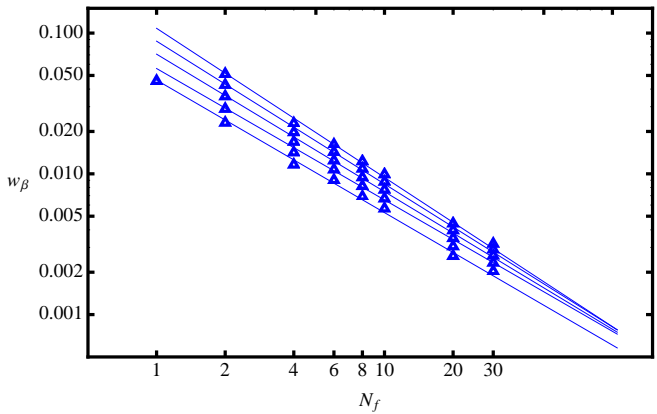}
\end{center}
\caption{Double-log-plot of the width $w_\beta(\Nf)$
  (cf. \Eqref{eq:wbeta}) of the Ginzburg region close to the finite
  temperature phase transition as a function of the flavor number
  $\Nf$. The tolerance criterion increases from bottom to top.
  The lowest data set corresponds to $2\%$, while the topmost
  corresponds to the $6\%$ criterion.
  Within the Ginzburg region, $2d$ Ising scaling can be
  observed. The Ginzburg region is suppressed for larger flavor number
  with a powerlaw $\sim 1/\Nf^{x_{\text{G}}}$ with varying from $x_{\text{G}}\simeq 0.93$ to $x_{\text{G}}\simeq 1.06$ (blue lines). }
\label{fig:Ginzburg}
\end{figure}
%----------------------------------------------------------
%
%Slight variations of the above-mentioned strategy (e.g., by modifying
%the 3\%\ criterion) lead to variations of the fitting exponent on the
%few percent level. 
Technically, the width becomes more difficult to
determine for increasing $\Nf$, as the resolution along the
temperature axis has to be finely adapted. {In Fig.~\ref{fig:Ginzburg}
we show our data in a range from $\Nf=1$ to $\Nf=30$.

Comparing our result for the scaling of the Ginzburg region,
$x_{\text{G}}\simeq 1$ with that of \cite{Kogut:1998ri} for
the scaling of the condensate value at the boundary of the mean-field
{region $x_{\text{MF}}\simeq 0.51$, the two different} values can be
connected by a plausibility argument: our criterion derives from a
quantity scaling with temperature, cf. \Eqref{eq:wbeta}, whereas
\cite{Kogut:1998ri} studied the scaling of the condensate in the
mean-field region. Since mean-field theory predicts a relation between
both with an exponent $\beta_{\text{MF}}=1/2$, it is reassuring that
the difference between \cite{Kogut:1998ri} for the scaling exponent of
the condensate value at the boundary of the mean-field region
$x_{\text{MF}}$ and the scaling exponent of the Ginzburg region
$x_{\text{G}}$ are separated by an amount on the order of
$\beta_{\text{MF}}$. A formal large-$\Nf$ argument
  \cite{Caracciolo:2005zu} also points into the same direction.

Let us conclude this part with a word of caution: Even though the
properties within the critical Ginzburg region are universal, the
boundary of this region measured in terms of $T_{\text{G}}$ as
  well as the boundary of the mean-field region may not exhibit the
same degree of universality.  This is, because $T_{\text{G}}$ is
determined by the temperature where the shorter-range thermal fermion
and higher-Matsubara boson fluctuations kick in.  Such sources of
non-universality should even be more pronounced for the location of
the boundary of the mean-field region and the corresponding exponent
$x_{\text{MF}}$, since the mean-field region is even further apart
from the universal critical regime. In fact, we observe an even
stronger definition dependence of our results for $x_{\text{MF}}$
which nevertheless is roughly of the same size as the lattice value
\cite{Kogut:1998ri}.

%
%--------------------------Figure--------------------------(classical regime)
\begin{figure}[t]
\begin{center}
\includegraphics[width=0.96\linewidth]{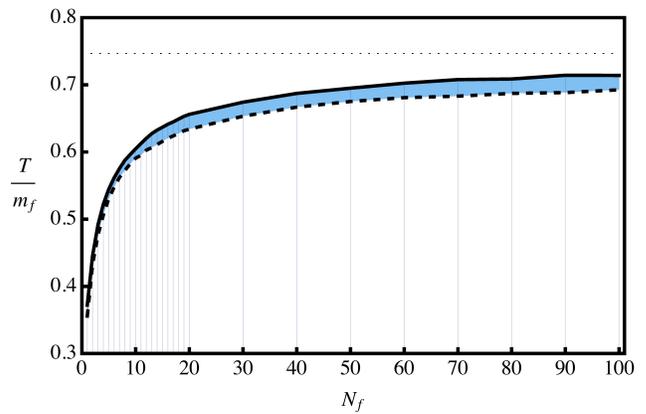}
\end{center}
\caption{Plot of the classical regime (dashed black) below the phase boundary in the many-flavor phase diagram. The critical region begins where the renormalized thermal mass of the bosonic fluctuations equals temperature, $T^{\text{cl}} = m(T^{\text{cl}})$. The physics in this region is dominated by the bosonic Matsubara zero mode.}
\label{fig:classical}
\end{figure}
%----------------------------------------------------------
%

Finally, let us analyse the size of the so-called {\it classical
  regime}, where the dimensional reduction, i.e. the dominance of the
bosonic zero mode sets in in the broken phase. Following
\cite{Kogut:1998ri}, we define the onset of the classical regime by
the value $T^{\text{cl}}$ of the temperature where this temperature
equals the renormalized bosonic mass $T^{\text{cl}} = m
(T^{\text{cl}})$. In the region $T^{\text{cl}}<T<\Tc$, the bosonic
Matsubara zero mode contributes significantly to the flow. This
classical region is shown as a shaded area in
Fig.~\ref{fig:classical}. Its width appears to be {approximately independent} of the flavor
number. {More quantitatively, we find that the width $w_{\text{cl}}(\Nf)$ of this region, measured in
units of the zero-temperature fermion mass, is }
\begin{equation}
w_{\text{cl}}(\Nf) =\frac{\Tc(\Nf) - T^{\text{cl}}(\Nf)}{\mf}\simeq\mathrm{const.}\,,
\end{equation}
aquires values in the range $w_{\text{cl}} \simeq 0.15 \dots 0.25$
with the smaller values holding for $\Nf\lesssim 10$. {Therefore,
  the classical regime remains finite even for $\Nf\to \infty$. This
  simply reflects the fact that the massless bosonic Matsubara zero
  mode will always exist near the phase transition even
  in the presence of an arbitrarily large but finite number of (thermally massive) fermion
  modes. This phenomenon remains completely invisible in the strict
  large-$\Nf$ limit. This observation indicates that the formal large-$\Nf$ limit has to be considered
  with some care: taking the limit $\Nf\to\infty$ on the level of the path integral does not appear to be equivalent 
  for all physical observables to solving the theory including bosonic fluctuations and then taking the limit~$\Nf\to\infty$}.

%----------------------------------------------------------------
\section{Conclusions}
\label{sec:conc}
%----------------------------------------------------------------

We have investigated the many-flavor phase diagram of the
$(2+1)$-dimensional Gross-Neveu model at finite temperature. At
zero-temperature, the model undergoes a quantum phase transition as a
function of the coupling strength, separating a massless from a
massive phase. The critical coupling marks a non-Gau\ss ian fixed
point which renders the model asymptotically safe, i.e. UV complete
despite its perturbative non-renormalizability. As this fixed point
has only one relevant direction, all long range observables can in
principle be predicted once a single physical scale is fixed. 
Hence, also the many-flavor finite-temperature phase diagram for the
massive model is also a pure prediction of the theory, if expressed,
for instance, in units of the vacuum fermion mass $\mf$. As mass
generation as well as symmetry restoration at finite temperature is a
non-perturbative process, we have used the functional RG for a
determination of various aspects of the phase diagram. In a 
conceptually comparatively simple framework, the functional RG gives
direct access to universal quantities in the critical region near the
phase transition as well as non-universal properties. In particular,
our predictions for the finite-temperature many-flavor phase boundary can be tested
by other non-perturbative approaches, such as lattice Monte-Carlo simulations.

In addition to the phase boundary, we have particularly concentrated
on the flavor-number dependence of the phase-diagram regions of (i)
local order, (ii) the Ginzburg critical region and (iii) the classical
region, all of which correspond to beyond-mean-field or finite-$\Nf$
phenomena.

The region of local (or short-range) order at temperatures right above
the phase transition characterizes temperature where a finite
condensate can be found on intermediate length scales. In addition to
possible influences on the spectral function, an understanding of this
region is particularly important for finite-volume studies, since
local order can easily be confused with long-range order for too small
volumes. It would be indeed  interesting to further analyze this question, 
e.~g., by directly employing lattice Monte-Carlo simulations.

The Ginzburg critical region corresponding to the range of
temperatures where the phase transition shows universal $2d$ Ising
scaling is especially relevant for an understanding of the approach to
the large-$\Nf$ limit. The non-analytic jump of the critical exponents
from $2d$ Ising to mean field at $\Nf \to \infty$ is realized by the
size of the Ginzburg critical region vanishing with
$1/\Nf^{x_{\text{G}}}$. Our result $x_{\text{G}}\simeq 1$ is in
accordance with other values in the literature~\cite{Kogut:1998ri,Caracciolo:2005zu}.

By contrast, we find that the classical regime, where the dynamics of
the bosonic Matsubara zero-mode sets in below $\Tc$, remains finite at
any $\Nf$. The existence of this mode cannot be inhibited by a
numerical dominance of a large number of flavors. The formal
large-$\Nf$ limit should therefore be viewed as a pure projection
technique onto the fermionic fluctuations rather than a means for
quantifying the dynamical prevalence of fluctuations. 

With respect to the applications of the Gross-Neveu model in
condensed-matter systems, the $\Nf=2$ model is phenomenologically most
important. Whereas some quantities such as the many-flavor phase
boundary may semi-quantitatively be estimated by higher-order studies
in a large-$\Nf$ expansion even down to smaller values of $\Nf$,
others such as the region of local order show strong variations
particularly near $\Nf=2$. This inhibits a general statement about the
in-principle applicability or inapplicability of large-$\Nf$
expansions for physical phenomena such as instabilities in graphite or
graphene, or secondary $d$- to $d+is$-wave pairing transitions in
nodal $d$-wave superconductors.

\acknowledgments The authors thank L.~Janssen, F.~Karbstein,
D.~Mesterhazy and M.~M.~Scherer for useful discussions.  JB
acknowledges support by the Helmholtz International Center for FAIR
within the LOEWE program of the State of Hesse.  HG acknowledges
support by the DFG under grants Gi~328/5-2 (Heisenberg
program). Moreover the authors acknowledge support by the research
training group ``Quantum and Gravitational Field'' (GRK 1523/1) and
the research training group FOR 723.

\appendix 

%----------------------------------------------------------------
\section{Threshold functions}\label{App:tf}
%----------------------------------------------------------------

The regulator dependence of the flow equations is encoded in dimensionless threshold
functions which are associated with 1PI diagrams incorporating bosonic and fermionic fields.
In this work we have employed a covariant linear
regulator~\cite{Litim:2000ci} for which the threshold functions at zero temperature are known
in closed form, see, e.g., Ref.~\cite{Hofling:2002hj}.

At finite temperature, the loop momentum integrals have to be
performed by replacing the time-like momentum component by the
corresponding fermionic or bosonic Matsubara frequencies. The
resulting threshold functions thus acquire a dependence on the
dimensionless temperature parameter $\tau=2\pi T/k$. For the treatment
of loop diagrams with mixed fermionic and bosonic propagators in the
loop, we follow the prescription of \cite{Braun:2008pi} in order to
achieve consistency with the derivative expansion. With the same
techniques and definitions as for $T=0$ and with
$v_d^{-1}=2^{d+1}\pi^{d/2}\Gamma(d/2)$ accounting for Euclidean volume
factors, we obtain the following finite temperature threshold functions
\begin{widetext}
\be 
%------------------------------------------------------------------------------------------------------
l_{n}^{d}(\tau,\omega_{B};\eta_{\sigma}) & = &
\frac{\delta_{n,0}+n}{2}\frac{v_{d-1}}{v_{d}}\frac{\tau}{2\pi}\sum_{n\in\mathds{Z}}
\int_{0}^{\infty}\mathrm{d}y\,y^{\frac{d-3}{2}} %\times \nn \\ 
%& &
((\tau c_{B,n})^{2}+y)\left(\mathcal{G}_{\sigma}(\tau
c_{B,n},y;\omega_{B})\right)^{n+1}\frac{\partial_{t}(Z_{\sigma,k}\,r_{B})}{Z_{\sigma,k}},\nn
%------------------------------------------------------------------------------------------------------
\\
%------------------------------------------------------------------------------------------------------ 
l_{n}^{(F)d}(\tau,\omega_{F};\eta_{\psi}) & = &
(\delta_{n,0}+n)\frac{v_{d-1}}{v_{d}}\frac{\tau}{2\pi}\sum_{n\in\mathds{Z}}
\int_{0}^{\infty}\mathrm{d}y\,y^{\frac{d-3}{2}} %\times \nn \\ & &
((\tau c_{F,n})^{2}+y)\left(\mathcal{G}_{\psi}(\tau
c_{F,n},y;\omega_{F})\right)^{n+1}\times \nn \\ 
& &  \frac{(1+r_{F})\partial_{t}(Z_{\sigma,k}\,r_{F})}{Z_{\psi,k}},\nn
%------------------------------------------------------------------------------------------------------
\\
%------------------------------------------------------------------------------------------------------
 l_{n,m}^{(FB)}(\tau,\omega_{F},\omega_{B};\eta_{\psi},\eta_{\sigma})
& = &
-\frac{1}{2}\frac{v_{d-1}}{v_{d}}\frac{\tau}{2\pi}\sum_{n\in\mathds{Z}}
\int_{0}^{\infty}\mathrm{d}y\,y^{\frac{d-3}{2}}\tilde{\partial}_{t}\left\{(\mathcal{G}_{\psi}(\tau
c_{F,n},y;\omega_{F}))^{n}(\mathcal{G}_{\sigma}(\tau
c_{F,n},y;\omega_{B}))^{m} \right\}, \nn
%------------------------------------------------------------------------------------------------------
\\ 
%------------------------------------------------------------------------------------------------------
m_{2}^{(F)d}(\tau,\omega_{F};\eta_{\psi}) & = &
-\frac{1}{2}\frac{v_{d-1}}{v_{d}}\frac{d}{d-1}\frac{\tau}{2\pi}\sum_{n\in\mathds{Z}}
\int_{0}^{\infty}\mathrm{d}y\,y^{\frac{d-1}{2}}\tilde{\partial}_{t}\left(\frac{\partial}{\partial
  y}\mathcal{G}_{\psi}(\tau c_{F,n},y;\omega_{F})\right)^{2}, \nn
%------------------------------------------------------------------------------------------------------
\\
%------------------------------------------------------------------------------------------------------
 m_{4}^{(F)d}(\tau,\omega_{F};\eta_{\psi}) & = &
-\frac{1}{2}\frac{v_{d-1}}{v_{d}}\frac{d}{d-1}\frac{\tau}{2\pi}\sum_{n\in\mathds{Z}}
\int_{0}^{\infty}\mathrm{d}y\,y^{\frac{d-1}{2}} %\times \nn \\ & &
((\tau
c_{F,n})^{2}+y)\tilde{\partial}_{t}\left(\frac{\partial}{\partial
  y}\left[(1+r_{F})\mathcal{G}_{\psi}(\tau
  c_{F,n},y;\omega_{F})\right]\right)^{2}, \nn
%------------------------------------------------------------------------------------------------------
\\ 
%------------------------------------------------------------------------------------------------------
m^{d}_{n,m}(\tau,\omega_{B,1},\omega_{B,2};\eta_{\sigma}) & = &
-\frac{1}{2}\frac{v_{d-1}}{v_{d}}\frac{d}{d-1}\frac{\tau}{2\pi}\sum_{n\in\mathds{Z}}
\int_{0}^{\infty}\mathrm{d}y\,y^{\frac{d-1}{2}} %\times \nn \\ & &
\tilde{\partial}_{t}\Biggl\{\left(\frac{\partial}{\partial y}\mathcal{G}_{\sigma}(\tau
c_{B,n},y;\omega_{B,1})\right)^{n/2}\times \nn \\
& &  \left(\frac{\partial}{\partial y}\mathcal{G}_{\sigma}(\tau c_{B,n},y;\omega_{B,2})\right)^{m/2}\Biggr\}, \nn
%------------------------------------------------------------------------------------------------------
\\
%------------------------------------------------------------------------------------------------------
m_{1,2}^{(FB)d}(\tau,\omega_{F},\omega_{B};\eta_{\psi},\eta_{\sigma})
& = &
-\frac{1}{2}\frac{v_{d-1}}{v_{d}}\frac{d}{d-1}\frac{\tau}{2\pi}\sum_{n\in\mathds{Z}}
\int_{0}^{\infty}\mathrm{d}y\,y^{\frac{d-1}{2}} %\times \nn \\ & &
\tilde{\partial}_{t}\left\{(1+r_{F})(\mathcal{G}_{\psi}(y;\omega_{F}))\left(\frac{\partial}{\partial
  y}\mathcal{G}_{\sigma}(y;\omega_{B})\right)\right\}. \nn
%------------------------------------------------------------------------------------------------------ 
\ee
where {the operator $\tilde{\partial}_t$ is} defined as 
\be 
\tilde{\partial}_{t}|_{\sigma} & = & \left[-2((\tau
c_n)^2+y)r_{B}'-\eta_{\sigma}r_{B}\right]\frac{\partial}{\partial
  r_{B}}, \\ 
\tilde{\partial}_{t}|_{\psi} & = & \left[-2((\tau
c_n)^2+y) r_{F}'-\eta_{\psi}r_{F}\right]\frac{\partial}{\partial r_{F}}, 
\ee
depending on whether it acts on bosonic or fermionic
propagators. Here, the Matsubara frequencies are encoded by the
variable $c_n$, acquiring the values $c_n=n$ for bosonic terms and
$c_n=n+\frac{1}{2}$ for fermionic terms.  ``Covariant'' optimized
shape functions now read
\be r_{B}((\tau c_{n})^{2}+y) & = & \left(\frac{1}{[(\tau
    c_{n})^{2}+y]}-1\right)\Theta(1-[(\tau c_{n})^{2}+y]),
\\ r_{F}((\tau c_{n})^{2}+y) & = & \left(\sqrt{\frac{1}{[(\tau
      c_{n})^{2}+y]}}-1\right)\Theta(1-[(\tau c_{n})^{2}+y]).  \ee
{The threshold functions for the linear regulator can be factorized in terms of
thermal and vacuum contributions.} In the
following, full thermal threshold functions and their vacuum
counterparts are distinguished by the dependence on the temperature
parameter $\tau$. We find
\be
%------------------------------------------------------------------------------------------------------  
l_{n}^{d}(\tau,\omega_{B};\eta_{\sigma}) & = &
\left(s_{0}^{d}(\tau)-\eta_{\sigma}\hat{s}_{0}^{d}(\tau)\right)l_{n}^{d}(\omega_{B};0),
\nn
%------------------------------------------------------------------------------------------------------  
\\
%------------------------------------------------------------------------------------------------------ 
l_{n}^{(F)d}(\tau,\omega_{F};\eta_{\psi}) & = &
\left(s_{0}^{(F)d}(\tau)-\eta_{\sigma}\hat{s}_{0}^{(F)d}(\tau)\right)l_{n}^{(F)d}(\omega_{F};0),
\nn
%------------------------------------------------------------------------------------------------------ 
\\ 
%------------------------------------------------------------------------------------------------------ 
l_{n,m}^{(FB)}(\tau,\omega_{F},\omega_{B};\eta_{\psi},\eta_{\sigma})
& = &
\left(\frac{1}{1+\omega_{F}}\right)^{n}\left(\frac{1}{1+\omega_{B}}\right)^{m}
\biggl\{n\left(s_{0}^{(F)d}(\tau)-\eta_{\psi}\hat{s}_{0}^{(F)d}(\tau)\right)l_{0}^{(F)d}(\omega_{F};0) + \nn \\
& &  m\left(s^{d}(\tau)-\eta_{\psi}\hat{s}^{d}(\tau)\right)l_{0}^{d}(\omega_{B};0)
\biggr\}, \nn 
%------------------------------------------------------------------------------------------------------ 
\\ 
%------------------------------------------------------------------------------------------------------ 
m_{2}^{(F)d}(\tau,\omega_{F};\eta_{\psi}) & = &
s_{0}^{(F)d}(\tau)m_{2}^{(F)d}(\omega_{F};0), \nn
%------------------------------------------------------------------------------------------------------ 
\\ 
%------------------------------------------------------------------------------------------------------ 
m_{4}^{(F)d}(\tau,\omega_{F};\eta_{\psi}) & = &
s_{0}^{(F)d}(\tau)\left(\frac{1}{1+\omega_{F}}\right)^{4} +
\frac{1-\eta_{\psi}}{2}t_{4}(\tau)\left(\frac{1}{1+\omega_{F}}\right)^{3}
- \nn \\ & &
\left(\frac{1-\eta_{\psi}}{4}t_{4}(\tau)+\frac{1}{4}s_{0}^{(F)d}(\tau)\right)\left(\frac{1}{1+\omega_{F}}\right)^{2},
\nn 
%------------------------------------------------------------------------------------------------------ 
\\
%------------------------------------------------------------------------------------------------------  
m^{d}_{n,m}(\tau,\omega_{B,1},\omega_{B,2};\eta_{\sigma}) & = &
s_{0}^{d}(\tau)m_{n,m}^{d}(\omega_{B,1},\omega_{B,2};0), \nn
\\ m_{1,2}^{(FB)d}(\tau,\omega_{F},\omega_{B};\eta_{\psi},\eta_{\sigma})
& = &
\left(s^{d}(\tau)-\eta_{\sigma}\hat{t}^{(FB)d}(\tau)\right)m_{1,2}^{(FB)d}(\omega_{F},\omega_{B};0,0). \nn
%------------------------------------------------------------------------------------------------------ 
\ee
The thermal threshold factors are themselves given by the following
Matsubara sums
\be
%------------------------------------------------------------------------------------------------------ 
s_{0}^{d}(\tau) & = &
\frac{v_{d-1}}{v_{d}}\frac{d}{d-1}\frac{\tau}{2\pi}\sum_{n\in\mathds{Z}}\Theta(1-c_{B,n}^{2}\tau^{2})\left(1-c_{B,n}^{2}\tau^{2}\right)^{\frac{d-1}{2}},
\nn
%------------------------------------------------------------------------------------------------------ 
\\
%------------------------------------------------------------------------------------------------------  
\hat{s}_{0}^{d}(\tau) & = &
\frac{v_{d-1}}{v_{d}}\frac{d}{d-1}\frac{\tau}{2\pi}\sum_{n\in\mathds{Z}}\Theta(1-c_{B,n}^{2}\tau^{2})\left(1-c_{B,n}^{2}\tau^{2}\right)^{\frac{d+1}{2}},
\nn 
%------------------------------------------------------------------------------------------------------ 
\\
%------------------------------------------------------------------------------------------------------ 
s_{0}^{(F)d}(\tau) & = &
\frac{v_{d-1}}{v_{d}}\frac{d}{d-1}\frac{\tau}{2\pi}\sum_{n\in\mathds{Z}}\Theta(1-c_{F,n}^{2}\tau^{2})\left(1-c_{F,n}^{2}\tau^{2}\right)^{\frac{d-1}{2}},
\nn 
%------------------------------------------------------------------------------------------------------ 
\\
%------------------------------------------------------------------------------------------------------ 
\hat{s}_{0}^{(F)d}(\tau) & = &
\frac{v_{d-1}}{v_{d}}\frac{d}{d-1}\frac{\tau}{2\pi}\sum_{n\in\mathds{Z}}\Theta(1-c_{F,n}^{2}\tau^{2})\left(1-c_{F,n}^{2}\tau^{2}\right)^{\frac{d-1}{2}}
%\times \nn \\ & & 
\left\{1-|c_{F,n}\tau|
_{2}F_{1}\left(-1/2,\frac{d-1}{2};\frac{d+1}{2};\frac{c_{F,n}^{2}\tau^{2}-1}{c_{F,n}^{2}\tau^{2}}\right)\right\}
\nn
%------------------------------------------------------------------------------------------------------  
\\
%------------------------------------------------------------------------------------------------------ 
s^{d}(\tau) & = &
\frac{v_{d-1}}{v_{d}}\frac{d}{d-1}\frac{\tau}{2\pi}\sum_{n\in\mathds{Z}}\Theta(1-c_{F,n}^{2}\tau^{2})\left(1-c_{F,n}^{2}\tau^{2}\right)^{\frac{d-1}{2}},
\nn 
%------------------------------------------------------------------------------------------------------ 
\\
%------------------------------------------------------------------------------------------------------ 
\hat{s}^{d}(\tau) & = &
\frac{v_{d-1}}{v_{d}}\frac{d}{d-1}\frac{\tau}{2\pi}\sum_{n\in\mathds{Z}}\Theta(1-c_{F,n}^{2}\tau^{2})\left(1-c_{F,n}^{2}\tau^{2}\right)^{\frac{d+1}{2}},
\nn 
%------------------------------------------------------------------------------------------------------ 
\\
%------------------------------------------------------------------------------------------------------ 
\hat{t}^{(FB)d}(\tau) & = &
\frac{v_{d-1}}{v_{d}}\frac{d}{d-1}\frac{\tau}{2\pi}\sum_{n\in\mathds{Z}}\Theta(1-c_{F,n}^{2}\tau^{2})\left(1-c_{F,n}^{2}\tau^{2}\right)^{\frac{d+1}{2}}
%\times \nn \\ & & 
|c_{F,n}\tau|^{-1}\,
_{2}F_{1}\left(1/2,\frac{d+1}{2};\frac{d+3}{2};\frac{c_{F,n}^{2}\tau^{2}-1}{c_{F,n}^{2}\tau^{2}}\right),
\nn 
%------------------------------------------------------------------------------------------------------ 
\\
%------------------------------------------------------------------------------------------------------ 
t_{4}(\tau) & = &
\frac{v_{d-1}}{v_{d}}\frac{d}{d-1}\frac{\tau}{2\pi}\sum_{n\in\mathds{Z}}\Theta(1-c_{F,n}^{2}\tau^{2})\left\{c_{F,n}^{2}\tau^{2}-1-2\ln(c_{F,n}\tau)\right\},
\nn 
%------------------------------------------------------------------------------------------------------ 
\ee
\end{widetext}
with the hypergeometric function $_{2}F_{1}(a,b;c;z)$~\cite{GR}.
\begin{center}
\begin{figure}[t!]
\begin{center}
\includegraphics[width=1\linewidth]{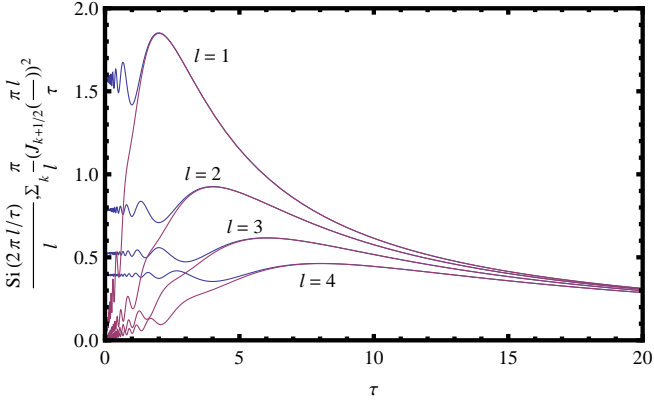}
\end{center}
\caption{Comparing $\frac{1}{l}\mathrm{Si}\left(\frac{2\pi
    l}{\tau}\right)$ (blue) and the approximant
  $\sum_{k=0}^{2}\frac{\pi}{l}\left(J_{k+1/2}\left(\frac{\pi
    l}{\tau}\right)\right)^{2}$ (magenta) as a function of $\tau$ for
  the Poisson indices $l=1,2,3,4$. The Poisson resummation over $l$
  can be done in closed form for the approximant. The deviation for
  $\tau\to0$ is immaterial since the sine integral enters the thermal
  threshold factor with a prefactor $\tau$ and the flow is stopped at
  finite $\tau$.
\label{fig:bessel}
}
\end{figure}
\end{center}
In the following, we give closed expressions for the thermal
threshold factors, specifically for three spacetime dimensions
($d=3$), which can be obtained by applying the Poisson resummation
formula
\be
\sum_{n\in\mathds{Z}}f(n)=\sum_{l\in\mathds{Z}}\int_{\mathds{R}}\,\mathrm{d}q\,f(q)\,\mathrm{e}^{-2\pi
  \mathrm{i}q l}, \ee
{and the identity~\cite{Lewin}}
\begin{equation}
 \Polylog{n}(-z)-(-1)^{n-1}\Polylog{n}(-1/z)=-\frac{(2\pi
  \mathrm{i})^{n}}{n!}B_{n}\left(\frac{\ln z}{2\pi
  \mathrm{i}}+\frac{1}{2}\right), 
\end{equation}
for $z\notin(0,1)$ where $B_{n}(z)$ are Bernoulli polynomials and
$\Polylog{n}(z)$ denotes the $n$th polylogarithm. With
\begin{eqnarray}
\ln\left(\mathrm{e}^{-2\pi\mathrm{i}/\tau}\right)&=&-2\pi\mathrm{i}+2\pi\mathrm{i}s_{F}(\tau),
\nonumber\\
\ln\left(-\mathrm{e}^{-2\pi\mathrm{i}/\tau}\right)&=&\mathrm{i}\pi-\frac{2\pi\mathrm{i}}{\tau}
+2\pi\mathrm{i}s_{B}(\tau),
\end{eqnarray}
and the definitions of the fermionic and bosonic floor
functions~\cite{Synatschke:2010ub} $s_{F}(\tau) =
\left\lfloor\frac{1}{\tau}+\frac{1}{2}\right\rfloor$, $s_{B}(\tau) =
\left\lfloor\frac{1}{\tau}\right\rfloor$, we arrive at
\begin{widetext}
\be s_{0}^{d}(\tau) & = &
\frac{d}{d-1}\frac{v_{d-1}}{v_{d}}\frac{1}{2\pi}\frac{1}{3}\left(-(\tau
+2\tau s_{B}(\tau))\left(-3+\tau^2 s_{B}(\tau)+\tau^2
s_{B}(\tau)^2\right)\right), \nn \\ \hat{s}_{0}^{d}(\tau) & = &
\frac{d}{d^{2}-1}\frac{v_{d-1}}{v_{d}}\frac{1}{2\pi}\frac{1}{15}\tau(1+2
s_{B}(\tau))\biggl(15-\tau^2 \left(10+\tau^2\right) s_{B}(\tau)+ \nn
\\ & & 2\tau^2 \left(-5+\tau^2\right) s_{B}(\tau)^2+6\tau^4
s_{B}(\tau)^3+3\tau^4 s_{B}(\tau)^4\biggr), \nn \\ s_{0}^{(F)d}(\tau)
& = & \frac{d}{d-1}\frac{v_{d-1}}{v_{d}}\frac{1}{2\pi}\frac{1}{6} \tau
s_{F}(\tau)\left(12 + (1 + 12 ) \tau^2 - 4 \tau^2
s_{F}(\tau)^2\right), \nn \\ \hat{s}_{0}^{(F)d}(\tau) & = &
\frac{d}{d-1}\frac{v_{d-1}}{v_{d}}\frac{1}{2\pi}\frac{1}{12} \biggl(4
s_{F}(\tau)+\left(4+\tau^2\right)s_{F}(\tau)-\tau^3 s_{F}(\tau)^2-4
\tau^2 s_{F}(\tau)^3 + \nn \\ & & 2 \tau^3 s_{F}(\tau)^4+\tau^2
\left(s_{F}(\tau) - \tau s_{F}(\tau)^2-4 s_{F}(\tau)^3+2 \tau
s_{F}(\tau)^4\right)\biggr), \nn \\ s^{d}(\tau) & = &
\frac{d}{d-1}\frac{v_{d-1}}{v_{d}}\frac{1}{2\pi}\frac{1}{6}\tau
s_{F}(\tau) \left(12+\tau^2-4 \tau^2 s_{F}(\tau)^2\right), \nn
\\ \hat{s}^{d}(\tau) & = &
\frac{d}{d^{2}-1}\frac{v_{d-1}}{v_{d}}\frac{1}{2\pi}\frac{1}{120}\left(
\tau s_{F}(\tau) (240 + 40 \tau^2 + 7 \tau^4 - 40 \tau^2 (4 + \tau^2)
s_{F}\tau^2 + 48 \tau^4 s_{F}\tau^4)\right), \nn
\\ \hat{t}^{(FB)d}(\tau) & = &
\frac{d}{d^{2}-1}\frac{v_{d-1}}{v_{d}}\frac{1}{2\pi} \frac{2}{3} \tau
s_ {F}(\tau) \left(4+\tau^2+\tau^2 s_{F}(\tau) (-\tau +2 s_{F}(\tau)
(-2+\tau s_{F}(\tau)))\right), \nn \\ t_{4}(\tau) & = &
\frac{d}{d-1}\frac{v_{d-1}}{v_{d}}\frac{1}{2\pi}\biggl(\frac{1}{6}
\tau s_{F}(\tau) (-12 - \tau^2 + 4 \tau^2 s_{F}(\tau)^2) + \frac{1}{3}
\tau (32 s_{F}(\tau) + 2 \tau (-12 - \tau^2) s_{F}(\tau)^2 + \nn \\ &
& 4 \tau^3 s_{F}(\tau)^4 - \frac{1}{20} \tau (-32 s_{F}(\tau) + 2 \tau
(40 + 20 \tau^2 + 7 \tau^4) s_{F}(\tau)^2 + \nn \\ & & 40 \tau^3 (-2 -
\tau^2) s_{F}(\tau)^4 + 32 \tau^5 s_{F}(\tau)^6)) + \dots\biggr). \nn
\ee
\end{widetext}
The expression for $t_{4}(\tau)$ is only {approximate since the Poisson
resummation could not be performed analytically.} However, after
performing the Poisson integration, the resulting epxression contains
a sine integral, which can be expanded in terms of Bessel functions,
$\mathrm{Si}(z)=\pi\sum_{k=0}^{\infty}\left(J_{k+1/2}\left(\frac{z}{2}\right)\right)^{2}$. Above,
we expanded to $k=2$ and cross-checked with numerical evaulation of
$t_{4}(\tau)$, see also Fig.~\ref{fig:bessel}. {The agreement is
satisfactory. In fact, we have checked that the difference between our approximate expression and the exact result does not 
lead to significant differences in our results for the phase diagram.}

\newpage  

%----------------------------------------------------------------

%

\end{document}